\RequirePackage{fix-cm}
\documentclass[preprint]{elsarticle}

\usepackage{geometry}
\usepackage[T1]{fontenc}
\usepackage{amsmath,amssymb,amsthm}
\usepackage{bm}
\usepackage{graphicx}
\usepackage{booktabs}
\usepackage{placeins}
\usepackage{siunitx}
\usepackage[hidelinks]{hyperref}
\usepackage{orcidlink}

\newcommand{\Weiss}{\mathrm{We}}
\newcommand{\Sc}{\mathrm{Sc}}
\newcommand{\Ma}{\mathrm{Ma}}
\newcommand{\Rey}{\mathrm{Re}}

\journal{Journal of Non-Newtonian Fluid Mechanics}

\begin{document}

\begin{frontmatter}
\title{A deterministic Fokker--Planck/lattice-Boltzmann micro--macro solver for dilute Hookean polymer solutions}

\author[a]{Xueqin Liu}
\author[a]{Shi Shu\,\orcidlink{0000-0002-5552-3840}}
\author[a]{Yaolong Yu}
\author[a,b,c]{Yuan Yu\corref{cor1}\,\orcidlink{0000-0001-5125-2492}}
\ead{yuyuan@xtu.edu.cn}
\author[a]{Hao Zhang}
\author[a]{Yuting Zhou}
\cortext[cor1]{Corresponding author.}
\address[a]{School of Mathematics and Computational Science, Xiangtan University, Xiangtan 411105, China}
\address[b]{National Center for Applied Mathematics in Hunan, Xiangtan 411105, China}
\address[c]{Hunan Key Laboratory for Computation and Simulation in Science and Engineering, Xiangtan University, Xiangtan 411105, China}

\begin{abstract}
Deterministic micro--macro simulation of polymer solutions requires the coupled evolution and spatial transport of a configuration distribution, together with the conversion of its moments into macroscopic stress and feedback to the flow. We develop a Fokker--Planck/lattice-Boltzmann solver for this coupling in two-dimensional dilute Hookean polymer solutions. The solver represents the distribution on the unbounded configuration space by Hermite coefficient fields, advances local configuration dynamics and physical-space transport in separate steps, and recovers Kramers stress from the second moment to achieve two-way coupling with a purified two-relaxation-time lattice-Boltzmann flow solver. To assess whether this configuration description recovers the corresponding macroscopic response, we use the exact second-moment closure of the continuous Hookean model as a macroscopic reference for comparison with analytical and independently discretized macroscopic solutions. The calculations reproduce velocity overshoot and damped oscillations in start-up Poiseuille flow, with velocity profiles approaching the analytical start-up transient under grid refinement; in four-roll flow with spatially nonuniform extension and stress feedback, the maximum full-domain relative differences in velocity and polymer stress are approximately $0.0309\%$ and $3.84\%$, respectively, over the tested Weissenberg numbers at corresponding sampling times. These comparisons support the solver's ability to recover the Hookean macroscopic response from configuration-distribution evolution. Further calculations of wall-bounded recirculation and open cross-slot flow exhibit the corresponding conformation responses and symmetric and asymmetric flow states, extending this deterministic micro--macro method to viscoelastic flows under different boundary constraints.
\end{abstract}


\end{frontmatter}

\section{Introduction}
\label{sec:introduction}

The viscoelastic behavior of dilute polymer solutions originates from the interaction between molecular configurations and the macroscopic flow. The flow stretches and orients polymer molecules, and the resulting nonequilibrium configuration distribution generates an additional stress that in turn modifies the flow field; molecular relaxation gradually returns the configurations toward equilibrium. Macroscopic constitutive models describe this process through the stress or conformation tensor, providing a concise route for computing viscoelastic flows. However, deriving a closed constitutive equation from a molecular model is not always feasible. Le Bris and Lelièvre~\cite{LeBrisLelievre2012} discussed the connection between microscopic dynamics and macroscopic constitutive relations: for nonlinear spring models, the evolution of low-order moments generally depends on configuration information that is not fully represented by these moments, and a closure approximation is therefore required. The finitely extensible nonlinear elastic (FENE) model and its Peterlin closure constitute a representative example.

A micro--macro description that retains the configuration distribution provides another route for handling such closure problems. For a given molecular model, the Fokker--Planck (FP) equation describes the evolution of the configuration distribution, while the polymer stress obtained from statistical averages over the distribution provides coupling to the macroscopic momentum equation. The molecular force law can thus enter the stress calculation directly, without first being converted into a closed macroscopic constitutive relation. This provides the main motivation for studying FP-based computational methods. The corresponding numerical difficulty is that the distribution depends on both physical position and molecular configuration: local deformation and relaxation alter configurations, the flow transports these configurations to different regions, and the resulting stress in turn affects the flow field that transports them. The configuration-space solution, physical-space transport, and flow feedback must therefore be considered together.

Following this approach, the present work investigates deterministic coupling between FP and the lattice Boltzmann method (LBM), using the Hookean dumbbell model initially to construct and evaluate the solver. For the Hookean model, the second moment closes exactly as the Oldroyd--B conformation equation, and directly solving this macroscopic equation is already sufficient to obtain the corresponding velocity, conformation, and stress. Accordingly, the significance of using FP for this model lies in developing a distribution-based computational method: we retain the evolution and transport of the configuration distribution to establish a micro--macro coupling implementation and test whether it recovers the known macroscopic response. The exact closure relation of the Hookean model provides an unambiguous reference for this test and prevents an additional constitutive closure approximation from being confounded with the evaluation of the method. With the same model parameters, physical-space diffusion, initial and boundary conditions, and corresponding sampling times, analytical solutions and independently discretized Oldroyd--B solutions can be used to assess the numerical behavior of the coupled implementation.

Deterministic configuration solvers have been implemented in several ways. Lozinski and Chauvière~\cite{LozinskiChauviere2003} used spectral and spectral-element methods to solve the two-dimensional FENE model, whereas Bao et al.~\cite{BaoLiuWang2025} coupled a deterministic variational particle method with a finite-element flow solver. These methods organize configuration evolution and stress calculation through different representations of the distribution. For the Hookean model considered here, the unbounded configuration space and Gaussian equilibrium distribution make Hermite functions a natural choice of basis. A Hermite expansion requires no finite artificial boundary in configuration space and represents the distribution by a set of coefficients that vary with physical position and time, allowing these coefficients to be updated locally and transported in physical space.

The relationship between Hermite coefficients and statistical moments also facilitates micro--macro coupling. Hetland et al.~\cite{Hetland2023} investigated the relationship among Hermite coefficients, Gauss--Hermite quadrature, and the stress of linearly elastic dumbbells in prescribed homogeneous flows. For nonuniform flows, Mizerová and She~\cite{MizerovaShe2018} combined a Hermite spectral discretization, Lagrange--Galerkin transport, and a Navier--Stokes solver. Beddrich et al.~\cite{Beddrich2024Fractional,Beddrich2025Turbulent} further studied polymeric flows with memory effects and discussed stress recovery under a particular Hermite scaling and the numerical convergence of the fully coupled system. These studies show how Hermite expansions connect local configuration dynamics, spatial transport, and macroscopic stress. Building on this foundation, the present work focuses on its connection to an LBM flow solver and on the behavior of this specific coupling strategy under different flow conditions.

In LBM-based viscoelastic computations, the configuration and flow problems can be assigned different roles. Ammar~\cite{Ammar2010} and Singh et al.~\cite{SinghSubramanianAnsumali2011} used LBM to solve the FP equation in configuration space; Bergamasco et al.~\cite{Bergamasco2013} coupled this type of configuration solver to a macroscopic finite-volume method and transported the polymer stress through physical space. Zou et al.~\cite{Zou2014} adopted another division of labor, in which LBM advances the macroscopic flow and a finite-volume method solves the constitutive equation. These methods differ not only in their discretization schemes, but also in the variables transported through physical space and in how stress feedback is obtained from those variables. In the present method, Hermite coefficient fields carry the configuration evolution and spatial transport, their statistical moments provide the stress, and LBM advances the macroscopic flow. This division requires coefficient transport, moment recovery, and force feedback to remain compatible at the discrete level, while the boundary data and update sequence must be specified for each geometry.

To realize this coupling, we construct a deterministic FP/LBM solver for two-dimensional dilute Hookean polymer solutions. Local deformation and relaxation act on the Hermite coefficients through an implicit configuration update, physical-space transport then advances the coefficient fields, and the recovered second moment gives the polymer stress through the Kramers relation. The stress divergence acts as the polymer force and is bidirectionally coupled to the purified two-relaxation-time lattice Boltzmann (P-TRT LBM) flow solver proposed by Yu et al.~\cite{Yu2026PTRT}. At walls and open boundaries, data for the flow and configuration variables must also be treated separately. In their study of the conformation tensor, Yu et al.~\cite{Yu2026ImprovedLBM} discussed the importance of solid-wall discretization for polymer variables; here, we prescribe treatments for the transported Hermite coefficient fields at solid walls, inlets, and outlets, and match the inlet coefficients to the required conformation moments. Together, these operations constitute the solver, while the specific feedback sequence and boundary implementation for each case are determined by the corresponding discrete conditions.

The numerical evaluation begins with flows for which clear macroscopic references are available. For start-up Poiseuille flow, the analytical solution of Waters and King~\cite{WatersKing1970} is used to examine velocity overshoot, damped oscillations, and the transient approximation under grid refinement. The periodic four-roll flow then extends the evaluation to spatially nonuniform extension and bidirectional stress feedback; the stress structures near stagnation points studied by Thomases and Shelley~\cite{ThomasesShelley2007} illustrate the demands that this flow places on conformation and stress calculations. Using the same macroscopic conditions and corresponding sampling times, we compare the FP results over the entire domain with results from an independently discretized Oldroyd--B model, considering the velocity, conformation, and polymer stress simultaneously. The former case tests the transient response in a controlled shear flow, whereas the latter examines the macroscopic fields under the combined effects of nonuniform transport and stress feedback.

Following these two types of reference comparison, the lid-driven cavity and open cross-slot flow are used to examine computational behavior under boundary constraints. The moving wall and closed recirculation in the cavity subject configurations to continuously varying local deformation, whereas the cross-slot combines inlet-configuration transport, central extension, and outlet flow. Studies by Arratia et al.~\cite{Arratia2006CrossChannel} and Poole et al.~\cite{PooleAlvesOliveira2007} show that a symmetric cross-slot geometry can develop an asymmetric viscoelastic response. We use this open flow to examine the computed symmetric and asymmetric states and use repeated computations and comparisons over a common time window to delimit the scope of the results. The cavity and cross-slot serve to demonstrate boundary responses and, together with the preceding analytical solution and independent macroscopic-field comparisons, form the evaluation of the solver.

The computations use a Hermite expansion with truncation parameter $M=2$ and fixed basis scaling. The evaluation focuses on the macroscopic response obtained from the evolution of the configuration coefficients and on computational behavior under the prescribed boundary conditions. Convergence with configuration resolution for the full distribution, nonnegativity of the reconstructed density, and extension to non-Hookean models require further study. Section~\ref{sec:model} introduces the mathematical model, Section~\ref{sec:method} presents the configuration discretization, spatial transport, and flow-coupling method, Section~\ref{sec:results} discusses the numerical results for the four cases, and Section~\ref{sec:conclusions} summarizes the principal findings and their scope.

\section{Mathematical model}
\label{sec:model}

We consider a two-dimensional, isothermal, incompressible dilute solution of Hookean dumbbells in a Newtonian solvent. The solvent contributes viscous stress, and the dumbbell configuration distribution determines the polymer stress. Let $\nu_0=\nu_s+\nu_p$ be the total zero-shear kinematic viscosity, with solvent and polymer contributions $\nu_s$ and $\nu_p$, and let $\lambda$ be the Hookean relaxation time. Transport of the distribution in physical space is regularized by an artificial diffusivity $K$.

For reference length $L_c$ and velocity $U_c$, the dimensionless groups are
\begin{equation}
 \begin{aligned}
  \Rey&=\frac{U_cL_c}{\nu_0},
  &\Weiss&=\frac{\lambda U_c}{L_c},\\
  \beta&=\frac{\nu_s}{\nu_0},
  &\Sc&=\frac{\nu_s}{K},
 \end{aligned}
 \label{eq:model-groups}
\end{equation}
Here $\Rey$ and $\Weiss$ measure inertia relative to viscous transport and polymer relaxation relative to advection, respectively, $\beta$ is the solvent viscosity fraction, and $\Sc$ controls the artificial diffusion. The geometry and driving of each case set $L_c$ and $U_c$; omitting artificial diffusion corresponds to $K=0$ and $\Sc=\infty$. 

The configuration scale is chosen so that the equilibrium probability density is $\psi_{\mathrm{eq}}=(2\pi)^{-1}\exp(-|\boldsymbol{Q}|^2/2)$ and its conformation tensor is $\boldsymbol{I}$. We scale $\boldsymbol{x}$, $t$, and $\boldsymbol{u}$ by $L_c$, $L_c/U_c$, and $U_c$, respectively; pressure divided by the constant reference density and kinematic stress both have scale $U_c^2$, while force per unit mass has scale $U_c^2/L_c$. In these variables, the coupled equations read
\begin{align}
 \begin{aligned}
  \partial_t\boldsymbol{u}
  +\boldsymbol{u}\mathbin{\cdot}\nabla\boldsymbol{u}
  &=-\nabla p_f+\frac{\beta}{\Rey}\Delta\boldsymbol{u}
    +\nabla\mathbin{\cdot}\boldsymbol{\tau}_p+\boldsymbol{f}_e,\\
  \nabla\mathbin{\cdot}\boldsymbol{u}&=0,
 \end{aligned}
 \label{eq:model-ns-nondimensional}\\
 \begin{aligned}
  \partial_t\psi+\boldsymbol{u}\mathbin{\cdot}\nabla\psi
  +\nabla_{\boldsymbol{Q}}\mathbin{\cdot}
   \left[(\nabla\boldsymbol{u})\boldsymbol{Q}\,\psi\right]
  &=\frac{1}{2\Weiss}\nabla_{\boldsymbol{Q}}\mathbin{\cdot}
    \left(\boldsymbol{Q}\psi+\nabla_{\boldsymbol{Q}}\psi\right)\\
  &\quad+\frac{\beta}{\Rey\,\Sc}\Delta\psi .
 \end{aligned}
 \label{eq:model-fp-nondimensional}
\end{align}
The unknowns are velocity $\boldsymbol{u}(\boldsymbol{x},t)$, normalized pressure $p_f(\boldsymbol{x},t)$, and the conditional probability density of the configuration, $\psi(\boldsymbol{x},\boldsymbol{Q},t)$, with $\boldsymbol{Q}\in\mathbb{R}^2$; $\boldsymbol{f}_e$ is the imposed force per unit mass. Operators $\nabla$ and $\Delta$ act in physical space unless a configuration-space subscript is shown, and $(\nabla\boldsymbol{u})_{\mu\xi}=\partial_\xi u_\mu$.

In eq.~\eqref{eq:model-fp-nondimensional}, advection transports the distribution with the flow, while the velocity gradient deforms it in configuration space. Hookean relaxation and Brownian diffusion act in configuration space, and the final term regularizes physical-space transport. The conditional density is normalized at each physical location:
\begin{equation}
 \int_{\mathbb{R}^{2}}\psi(\boldsymbol{x},\boldsymbol{Q},t)
 \,\mathrm{d}\boldsymbol{Q}=1,
 \qquad
 \psi\rightarrow0\quad\text{as }|\boldsymbol{Q}|\rightarrow\infty.
 \label{eq:model-normalization}
\end{equation}
We assume finite second moments and sufficient decay of the configuration-space flux for the boundary terms in moment integration to vanish.

The second moment supplies the stress feedback through the Hookean Kramers relation described by Bird et al.~\cite{Bird1987DPL2}:
\begin{equation}
 \boldsymbol{A}
 =\int_{\mathbb{R}^{2}}\boldsymbol{Q}\otimes\boldsymbol{Q}\,
   \psi\,\mathrm{d}\boldsymbol{Q},
 \qquad
 \boldsymbol{\tau}_p
 =\frac{1-\beta}{\Rey\,\Weiss}
  (\boldsymbol{A}-\boldsymbol{I}).
 \label{eq:model-kramers}
\end{equation}
Here $\boldsymbol{I}$ is the two-dimensional identity tensor. Taking the second moment of eq.~\eqref{eq:model-fp-nondimensional} then gives the diffusive Oldroyd--B conformation equation
\begin{equation}
 \partial_t\boldsymbol{A}
 +\boldsymbol{u}\mathbin{\cdot}\nabla\boldsymbol{A}
 - (\nabla\boldsymbol{u})\boldsymbol{A}
 - \boldsymbol{A}(\nabla\boldsymbol{u})^{\mathsf T}
 =-\frac{1}{\Weiss}(\boldsymbol{A}-\boldsymbol{I})
 +\frac{\beta}{\Rey\,\Sc}\Delta\boldsymbol{A}.
 \label{eq:model-oldroydb}
\end{equation}
This connects the constitutive model of Oldroyd~\cite{Oldroyd1950} to the micro--macro description reviewed by Le Bris and Lelièvre~\cite{LeBrisLelievre2012}; Dębiec and Süli~\cite{DebiecSuli2025} establish rigorous closure in the corresponding two-dimensional diffusive setting. The closed equation supplies the macroscopic reference for the present coefficient-based solver. With the same model parameters, physical-space diffusion, and initial and boundary data, the two continuous descriptions have the same velocity, conformation, and polymer stress. Their numerical realizations evolve different variables: the FP solver advances distribution coefficients and recovers $\boldsymbol{A}$ from their moments, while the reference solver advances $\boldsymbol{A}$ directly. Comparing the independently discretized solutions at the same sampling time for each field therefore tests the coupled transport and stress feedback.

\section{Numerical method}
\label{sec:method}

Hermite coefficient fields carry the configuration information through the calculation. Local deformation and relaxation update these coefficients within each cell, and spatial transport moves them through the flow domain. Their moments provide the Kramers stress, whose divergence enters the P-TRT flow update; the resulting velocity supplies the next configuration update. Sections~\ref{sec:method-hermite}--\ref{sec:method-ptrt} define these operations, and Sections~\ref{sec:method-boundaries} and~\ref{sec:method-coupling} specify the boundary data and exchange times.

Unless stated otherwise, the discrete equations use lattice units with $\Delta x=\Delta y=\Delta t=1$. The assignments $\nu_s=\beta U_cL_c/\Rey$, $\nu_p=(1-\beta)U_cL_c/\Rey$, $\lambda=\Weiss L_c/U_c$, and $K=\nu_s/\Sc$ convert eq.~\eqref{eq:model-groups} into lattice parameters. In particular, the local configuration update uses the lattice relaxation time $\lambda$.

\subsection{Configuration representation and local coefficient evolution}
\label{sec:method-hermite}

We represent the decaying configuration density on its unbounded domain using scaled physicists' Hermite functions, following Fok et al.~\cite{FokGuoTang2002} and Beddrich et al.~\cite{Beddrich2025Turbulent}. With modal indices $a$ and $b$, a fixed scaling parameter $\alpha$, and one-dimensional Hermite polynomials $H_a$, the basis, expansion, and dual projection are
\begin{equation}
\begin{gathered}
 \widetilde H_a(q)=\frac{H_a(\alpha q)}{\sqrt{2^a a!}}
 e^{-\alpha^2q^2},
 \qquad
 \psi_M=\sum_{a=0}^{M}\sum_{b=0}^{M}
 \widehat\psi_{ab}\,
 \widetilde H_a(Q_x)\widetilde H_b(Q_y),\\
 \widehat\psi_{ab}=\frac{\alpha^2}{\pi}
 \int_{\mathbb{R}^2}\psi
 \widetilde H_a(Q_x)\widetilde H_b(Q_y)
 e^{\alpha^2|\boldsymbol{Q}|^2}\,\mathrm{d}\boldsymbol{Q}.
\end{gathered}
\label{eq:method-hermite-expansion}
\end{equation}
The polynomial identities for this convention are given in \ref{sec:method-hermite-identities}. The baseline calculations use the square truncation $M=2$ with $\alpha=1/\sqrt2$, giving nine coefficient fields per cell. At equilibrium, only $\widehat\psi_{00}=1/(2\pi)$ is nonzero.

This scaling gives a direct relation between the coefficients and the moments needed for stress recovery. Writing $m_0=\int\psi_M\,\mathrm{d}\boldsymbol{Q}$ and $m_{ij}=\int Q_iQ_j\psi_M\,\mathrm{d}\boldsymbol{Q}$, with both integrals over $\mathbb{R}^2$, we obtain
\begin{equation}
 m_0=2\pi\widehat\psi_{00},
 \qquad
 \boldsymbol{m}=2\pi
 \begin{pmatrix}
 \widehat\psi_{00}+\sqrt2\,\widehat\psi_{20} & \widehat\psi_{11}\\
 \widehat\psi_{11} & \widehat\psi_{00}+\sqrt2\,\widehat\psi_{02}
 \end{pmatrix}.
 \label{eq:method-coefficient-moments}
\end{equation}
Thus, the zeroth mode and three second-order modes determine the local probability mass and raw second moments. These are the quantities passed from the configuration representation to the flow solver.

For a prescribed local velocity gradient, the configuration part of the FP equation becomes a linear system in the coefficients. Holding the gradient fixed over the substep and applying implicit Euler gives
\begin{equation}
\begin{split}
 \frac{\widehat\psi_{ab}^{\mathrm{loc}}-\widehat\psi_{ab}^{n}}{\Delta t}
 ={}&C_1^{ab}\widehat\psi_{ab}^{\mathrm{loc}}
 +C_2^{ab}\widehat\psi_{a-1,b-1}^{\mathrm{loc}}
 +C_3^{ab}\widehat\psi_{a-1,b+1}^{\mathrm{loc}}\\
 &+C_4^{ab}\widehat\psi_{a+1,b-1}^{\mathrm{loc}}
 +C_5^{ab}\widehat\psi_{a,b-2}^{\mathrm{loc}}
 +C_6^{ab}\widehat\psi_{a-2,b}^{\mathrm{loc}},
\end{split}
\label{eq:method-local-update}
\end{equation}
Coefficients outside $0\le a,b\le M$ are set to zero. At $\alpha^2=1/2$, the six couplings are
\begin{align}
 C_1^{ab}&=a\,\partial_x u_x+b\,\partial_y u_y-\frac{a+b}{2\lambda},
 &C_2^{ab}&=\sqrt{ab}\,(\partial_x u_y+\partial_y u_x),
 \label{eq:method-c12}\\
 C_3^{ab}&=\sqrt{a(b+1)}\,\partial_y u_x,
 &C_4^{ab}&=\sqrt{b(a+1)}\,\partial_x u_y,
 \label{eq:method-c34}\\
 C_5^{ab}&=\sqrt{b(b-1)}\,\partial_y u_y,
 &C_6^{ab}&=\sqrt{a(a-1)}\,\partial_x u_x.
 \label{eq:method-c56}
\end{align}
The gradient terms describe deformation, while $(a+b)/(2\lambda)$ governs relaxation of the nonzero modes. The $a=b=0$ row preserves $\widehat\psi_{00}$, and the equations for $\widehat\psi_{20}$, $\widehat\psi_{11}$, and $\widehat\psi_{02}$ involve only these four moment-carrying coefficients. Their local moment equation is derived in \ref{sec:method-hermite-identities}. This closed low-order structure motivates the baseline truncation; the macroscopic comparisons assess how its moments are transported and coupled to the flow.

The implementation solves the baseline $9\times9$ system in each cell. Four-roll and Poiseuille flow use pivoted dense elimination with damped Anderson-accelerated Jacobi (AAJ) as a fallback; cavity and cross-slot flow use damped AAJ directly. The zeroth coefficient is retained through the solve, with restoration to its incoming value after the Poiseuille, cavity, and cross-slot updates. Solver controls and stopping criteria are collected in \ref{sec:method-local-solvers}. The resulting coefficients enter physical-space transport.

\subsection{Physical-space transport}
\label{sec:method-transport}

After local configuration evolution, the coefficients undergo advection and artificial diffusion in physical space. Because the Hermite basis depends only on $\boldsymbol{Q}$, each coefficient satisfies
\begin{equation}
 \partial_t\widehat\psi_{ab}
 +\boldsymbol{u}\mathbin{\cdot}\nabla\widehat\psi_{ab}
 =K\Delta\widehat\psi_{ab}.
 \label{eq:method-physical-transport}
\end{equation}
We use finite-volume transport in all four flow configurations and an alternative scalar lattice-Boltzmann discretization in the Poiseuille comparison. The finite-volume formulation treats advection explicitly, with diffusion advanced either in a subsequent implicit step or within the explicit update, as specified below.

The advective face fluxes use the fifth-order WENO-Z reconstruction of Borges et al.~\cite{Borges2008WENOZ} with local Lax--Friedrichs face fluxes. Since the LBM velocity is weakly compressible, the explicit advection operator includes the correction
\begin{equation}
 -\nabla_h\mathbin{\cdot}(\boldsymbol{u}\widehat\psi_{ab})
 +\widehat\psi_{ab}\,\nabla_h\mathbin{\cdot}\boldsymbol{u}.
 \label{eq:method-nonconservative-correction}
\end{equation}
The velocity divergence in the correction uses the same face velocities as the advective fluxes. For a constant coefficient and matching boundary data, each face flux equals that constant times the face velocity, so the two discrete terms cancel.
A two-stage second-order Runge--Kutta update, following Shu and Osher~\cite{ShuOsher1988}, advances this operator with CFL substeps when required.

In the finite-volume Poiseuille branch and in four-roll and cavity flow, advection is followed by Crank--Nicolson alternating-direction implicit (CN--ADI) diffusion, following Peaceman and Rachford~\cite{PeacemanRachford1955}:
\begin{equation}
 \widehat\psi^{a}=\mathcal{A}_{\Delta t}\widehat\psi^{\mathrm{loc}},
 \qquad
 \widehat\psi^{n+1}=\mathcal{D}^{\mathrm{CN\text{-}ADI}}_{\Delta t}
 \widehat\psi^{a},
 \label{eq:method-poiseuille-transport}
\end{equation}
Here $\mathcal{A}_{\Delta t}$ denotes WENO-Z/LLF--RK2 advection and $\mathcal{D}^{\mathrm{CN\text{-}ADI}}_{\Delta t}$ denotes diffusion. For $K=0$, the diffusion operator is the identity and the ADI sweeps are omitted. The cross-slot mask instead enters through a discrete Laplacian in both Runge--Kutta right-hand sides, with substeps constrained by advection and explicit diffusion together.

The alternative Poiseuille branch uses a regularized scalar D2Q9 discretization of the full advection--diffusion equation, eq.~\eqref{eq:method-physical-transport}. On this lattice, $\boldsymbol{c}_i$ and $w_i$ are the velocities and weights, and $c_s^2=1/3$; the lattice definitions are given in \ref{sec:method-lattice-details}. For each coefficient, the scalar populations $h_i^{ab}$ satisfy
\begin{subequations}
\begin{align}
 h_i^{ab}(\boldsymbol{x}+\boldsymbol{c}_i\Delta x,t+\Delta t)
 &=h_i^{ab,\mathrm{eq}}+
 \left(1-\frac{1}{\tau_\phi}\right)h_i^{ab,\mathrm{neq,reg}}
 +w_i\Delta t\,S_{ab},\\
 h_i^{ab,\mathrm{eq}}
 &=w_i\widehat\psi_{ab}\left(1+
 \frac{\boldsymbol{c}_i\mathbin{\cdot}\boldsymbol{u}}{c_s^2}\right),
 \qquad \widehat\psi_{ab}=\sum_i h_i^{ab},\\
 h_i^{ab,\mathrm{neq,reg}}
 &=\frac{w_i}{c_s^2}\boldsymbol{c}_i\mathbin{\cdot}
 \left(\boldsymbol{J}_{ab}-\widehat\psi_{ab}\boldsymbol{u}\right),
 \qquad \boldsymbol{J}_{ab}=\sum_i\boldsymbol{c}_i h_i^{ab},\\
 S_{ab}&=\widehat\psi_{ab}\nabla_h\mathbin{\cdot}\boldsymbol{u},
 \qquad \tau_\phi=\frac12+\frac{K\Delta t}{c_s^2\Delta x^2}.
\end{align}
\label{eq:method-scalar-lbm-transport}
\end{subequations}
The source supplies the nonconservative correction, and regularization retains the first nonequilibrium moment following Latt and Chopard~\cite{LattChopard2006}. The FVM and LBM comparisons therefore differ in the transport of the FP coefficients, while sharing the local Hermite update, moment recovery, Kramers feedback, and P-TRT flow solver.

\subsection{Moment recovery and polymer force}
\label{sec:method-moments}

The transported coefficients determine the local probability mass and second moments through eq.~\eqref{eq:method-coefficient-moments}. In the implementation, these moments are evaluated by reconstructing the truncated density at tensor-product Gauss--Hermite nodes:
\begin{equation}
 \psi_{rs}=\sum_{a=0}^{2}\sum_{b=0}^{2}
 \widehat\psi_{ab}\widetilde H_a(q_r)\widetilde H_b(q_s),
 \qquad 0\le r,s\le2.
 \label{eq:method-reconstruction}
\end{equation}
For $M=2$ and $\alpha=1/\sqrt2$, the one-dimensional nodes and weights are
\begin{equation}
 q_r\in\{-\sqrt3,0,\sqrt3\},
 \qquad
 w_r=\frac{\sqrt{\pi}/\alpha}
 {3\,\widetilde H_2(q_r)^2\exp(\alpha^2q_r^2)},
 \qquad
 W_{rs}=w_rw_s .
 \label{eq:method-quadrature-rule}
\end{equation}
With $r$ and $s$ indexing the nodes, the tensor-product weights $W_{rs}$ give
\begin{equation}
\begin{gathered}
 m_0=\sum_{r,s=0}^{2}W_{rs}\psi_{rs},
 \qquad
 m_{xx}=\sum_{r,s=0}^{2}q_r^2W_{rs}\psi_{rs},\\
 m_{xy}=\sum_{r,s=0}^{2}q_rq_sW_{rs}\psi_{rs},
 \qquad
 m_{yy}=\sum_{r,s=0}^{2}q_s^2W_{rs}\psi_{rs}.
\end{gathered}
 \label{eq:method-quadrature-moments}
\end{equation}
For each coordinate, the polynomial degree in these weighted moment integrals is at most four. The three-point rule therefore integrates the required moments of $\psi_M$ exactly and recovers eq.~\eqref{eq:method-coefficient-moments}. The zeroth moment $m_0$ measures local probability normalization, and the raw second moments $m_{ij}$ supply the stress.

Poiseuille, four-roll, and cavity flow use the raw-second-moment Kramers map
\begin{equation}
 \tau^{\mathrm{P}}_{p,ij}=\frac{\nu_p}{\lambda}
 \left(m_{ij}-\delta_{ij}\right),
 \label{eq:method-stress-map-poiseuille}
\end{equation}
where $\delta_{ij}$ is the Kronecker delta and the stress is in lattice units. The zeroth moment is monitored separately; four-roll flow applies no normalization or projection after transport.

Cross-slot flow obtains the conformation and stress from normalized moments:
\begin{equation}
 A_{ij}=\frac{m_{ij}}{m_0},
 \qquad
 \tau_{p,ij}=\frac{\nu_p}{\lambda}
 \left(A_{ij}-\delta_{ij}\right),
 \label{eq:method-stress-map}
\end{equation}
The division by $m_0$ cancels a common amplitude factor in the local coefficient vector, making the recovered conformation and stress invariant under that rescaling. Probability-mass deviations remain available through the separate diagnostic $m_0$. The calculation is terminated if this moment is nonfinite or does not satisfy $m_0>10^{-12}$. At $m_0=1$, the two stress maps coincide with the normalized Hookean relation in eq.~\eqref{eq:model-kramers}.

The divergence of the recovered stress supplies the polymer force. Cross-slot flow uses the centred stencil
\begin{equation}
 \begin{aligned}
 F_{p,x}&=\frac{\tau_{p,xx}^{E}-\tau_{p,xx}^{W}}{2\Delta x}
          +\frac{\tau_{p,xy}^{N}-\tau_{p,xy}^{S}}{2\Delta y},\\
 F_{p,y}&=\frac{\tau_{p,yx}^{E}-\tau_{p,yx}^{W}}{2\Delta x}
          +\frac{\tau_{p,yy}^{N}-\tau_{p,yy}^{S}}{2\Delta y}.
 \end{aligned}
 \label{eq:method-stress-divergence}
\end{equation}
where $E$, $W$, $N$, and $S$ denote neighbouring grid locations. Poiseuille, four-roll, and cavity flow use the isotropic D2Q9 derivative operator. Periodic wrapping and wall or open-boundary values for these derivatives are specified in Section~\ref{sec:method-boundaries}.

\subsection{P-TRT solution of the flow equations}
\label{sec:method-ptrt}

The polymer force is incorporated into the P-TRT flow solver of Yu et al.~\cite{Yu2026PTRT}. This solver follows the regularized lattice-Boltzmann approach of Latt and Chopard~\cite{LattChopard2006} and Yu et al.~\cite{YuQinChen2025TRTRLB}, with separate relaxation of viscous and non-hydrodynamic modes. For the D2Q9 lattice, the collision equilibrium is
\begin{equation}
 f_i^{\mathrm{eq}}=w_i\rho\left[
 1+\frac{\boldsymbol{c}_i\mathbin{\cdot}\boldsymbol{u}}{c_s^2}
 +\frac{(\boldsymbol{c}_i\mathbin{\cdot}\boldsymbol{u})^2}{2c_s^4}
 -\frac{|\boldsymbol{u}|^2}{2c_s^2}
 +\frac{H^{(3)}_{i,\mu\nu\kappa}
 u_\mu u_\nu u_\kappa}{6c_s^6}
 \right],
 \label{eq:method-feq}
\end{equation}
Here $\rho$ is the lattice density, repeated Greek indices are summed over Cartesian components, and the third-order Hermite tensor is defined in \ref{sec:method-lattice-details}.

The force convention is introduced through the nonequilibrium populations and their moments. With $\bar i$ denoting the direction opposite to $i$, define
\begin{equation}
 \begin{aligned}
  \delta f_i&=f_i-f_i^{\mathrm{eq}},
  &a_\mu^{(1)}&=\sum_j c_{j\mu}\delta f_j,\\
  \Phi_i&=(c_{ix}^2-c_s^2)(c_{iy}^2-c_s^2),
  &S_G&=\sum_j\Phi_j\delta f_j .
 \end{aligned}
 \label{eq:method-ptrt-defs}
\end{equation}
The retained sectors and force projection are
\begin{align}
 R_i^{(1)}&=w_i\frac{\boldsymbol{c}_i\mathbin{\cdot}
                    \boldsymbol{a}^{(1)}}{c_s^2},
 &R_i^{(2)}&=\frac{\delta f_i+\delta f_{\bar i}}{2}
 -\frac{w_i\Phi_i}{4c_s^8}S_G,
 \label{eq:method-reg12}\\
 F_i&=w_i\frac{\boldsymbol{c}_i\mathbin{\cdot}\boldsymbol{F}}{c_s^2},
 &R_i^{(3)}&=\frac{\delta f_i-\delta f_{\bar i}}{2}
             +\frac{\Delta t}{2}F_i,
 \label{eq:method-reg3}\\
 \boldsymbol{F}&=\boldsymbol{F}_e+\boldsymbol{F}_p,
 &\boldsymbol{F}_p&=\nabla_h\mathbin{\cdot}\boldsymbol{\tau}_p .
 \label{eq:method-force-projection}
\end{align}
The total force $\boldsymbol{F}$ combines imposed and polymer contributions, normalized by the reference density $\rho_0=1$; thus $\boldsymbol{F}_e$ is the lattice counterpart of $\boldsymbol{f}_e$ in eq.~\eqref{eq:model-ns-nondimensional}. The first-order sector $R_i^{(1)}$ is treated separately, the even sector $R_i^{(2)}$ has its ghost contribution removed, and the odd sector $R_i^{(3)}$ includes the half-force shift. With relaxation rates $\omega_+=1/\tau_+$ and $\omega_-=1/\tau_-$, collision gives
\begin{equation}
\begin{split}
 f_i^{\mathrm{pc}}={}&f_i^{\mathrm{eq}}
 +(1-\omega_+)R_i^{(1)}
 +(1-\omega_+)R_i^{(2)}
 +(1-\omega_-)R_i^{(3)}\\
 &+\left(1-\frac{\omega_+}{2}\right)G_i\Delta t
  +\left(1-\frac{\omega_+}{2}\right)F_i\Delta t .
\end{split}
\label{eq:method-ptrt-update}
\end{equation}
The same projected force enters the odd-sector shift and the source term, fixing its contribution to the momentum update. The diagonal cubic-velocity correction is
\begin{equation}
 G_i=-\frac{w_i}{2c_s^4}\left[
 (c_{ix}^2-c_s^2)\,\partial_x(\rho u_x^3)
 +(c_{iy}^2-c_s^2)\,\partial_y(\rho u_y^3)
 \right],
 \label{eq:method-cubic-correction}
\end{equation}
Poiseuille, four-roll, and cavity flow retain this correction, while cross-slot flow sets $G_i=0$. The derivatives in the four-roll correction use periodic centred differences along the coordinate axes.

Macroscopic recovery uses the corresponding half-force term, while the relaxation times set the solvent viscosity and the odd-sector rate:
\begin{align}
 \rho&=\sum_i f_i,
 &\rho\boldsymbol{u}&=\sum_i\boldsymbol{c}_i f_i
 +\frac{\Delta t}{2}\boldsymbol{F},
 \label{eq:method-macroscopic}\\
 \nu_s&=c_s^2\left(\tau_+-\frac12\right)\Delta t,
 &\left(\tau_+-\frac12\right)
  \left(\tau_--\frac12\right)&=\frac{3}{16}.
 \label{eq:method-relaxation}
\end{align}
The force time level follows the update sequence in Section~\ref{sec:method-coupling}. The isothermal lattice pressure fluctuation is $p_{\mathrm{LB}}=c_s^2(\rho-\rho_0)$; division by $U_c^2$ gives $p_f$ up to a reference constant. In the low-Mach regime, density deviations scale as $O(\mathrm{Ma}_{\max}^2)$, where $\mathrm{Ma}_{\max}=\max_{\boldsymbol{x}}|\boldsymbol{u}|/c_s$, and the flow update approaches the incompressible momentum equation. Density variation, $\mathrm{Ma}_{\max}$, and flux balance are monitored following the LBM framework described by Krüger et al.~\cite{Kruger2017LBM}.

\subsection{Boundary data and discrete derivatives}
\label{sec:method-boundaries}

The local configuration update uses the velocity gradient, while stress feedback uses the divergence of the recovered stress. In Poiseuille, four-roll, and cavity flow, the interior velocity gradient is evaluated with the isotropic D2Q9 difference
\begin{equation}
 (\partial_\xi u_\mu)_h
 =c_s^{-2}\sum_{i=1}^{8}w_i c_{i\xi}
 u_\mu(\boldsymbol{x}+\boldsymbol{c}_i),
 \qquad \xi,\mu\in\{x,y\},
 \label{eq:method-gradient-poiseuille}
\end{equation}
Four-roll flow wraps the stencil periodically. At channel walls and all four cavity walls, the normal derivative uses a second-order one-sided difference. Cross-slot flow uses the centred coordinate differences
\begin{equation}
 (\partial_xu_\mu)_h=\frac{u_\mu^{E}-u_\mu^{W}}{2\Delta x},
 \qquad
 (\partial_yu_\mu)_h=\frac{u_\mu^{N}-u_\mu^{S}}{2\Delta y},
 \label{eq:method-gradient-crossslot}
\end{equation}
A fluid neighbour supplies its nodal velocity. A solid neighbour supplies $-\boldsymbol{u}_P$ to impose mid-link no slip; an exterior inlet neighbour supplies the prescribed velocity, and other open-boundary neighbours supply $\boldsymbol{u}_P$ for zero normal gradient.

Stress derivatives use boundary values of the stress itself. In cross-slot flow, fluid neighbours supply recovered stress, solid neighbours and outlets use local stress for zero normal gradient, and inlet neighbours supply the prescribed inlet stress constructed below. Poiseuille and cavity flow use the isotropic D2Q9 operator in the interior and second-order one-sided stress differences at walls; four-roll flow uses periodic wrapping.

At the cross-slot inlets, the coefficient and stress data are constructed from the same target conformation. Let $j_{\min}$ be the first cell-centre index across an arm, $j$ the transverse index, and $N_W$ the number of cells across its width $W=N_W\Delta y$. The parabolic profile and its signed dimensionless shear rate are
\begin{equation}
 \begin{aligned}
  \eta&=\frac{j-j_{\min}+1/2}{N_W},
  &\frac{u_x}{U_0}&=\chi_{\mathrm{in}}\,4\eta(1-\eta),\\
  \dot\gamma
  &=\frac{\partial(u_x/U_0)}{\partial(y/W)}
    =\chi_{\mathrm{in}}\,4(1-2\eta),\\
  \chi_{\mathrm{in}}
  &=\begin{cases}
     +1,&\text{left inlet},\\
     -1,&\text{right inlet}.
    \end{cases}
 \end{aligned}
 \label{eq:method-inlet-profile}
\end{equation}
Here $\eta\in(0,1)$ locates the cell centre across the arm, $U_0$ is the peak inlet speed, and $\chi_{\mathrm{in}}$ distinguishes the two inlet directions. The dimensional shear rate is $(U_0/W)\dot\gamma$. Using the local steady simple-shear relation of Bird et al.~\cite{Bird1987DPL2} and Oldroyd~\cite{Oldroyd1950}, we prescribe
\begin{equation}
 A_{11}=1+2\left[\Weiss\,\dot\gamma(y)\right]^2,
 \qquad
 A_{12}=\Weiss\,\dot\gamma(y),
 \qquad
 A_{22}=1.
 \label{eq:method-inlet-conformation}
\end{equation}
This construction sets the inlet data from the local shear rate, with $\Weiss=\lambda U_0/W$; physical-space diffusion is retained in the interior evolution. The signed shear sets the sign of the off-diagonal component. At $M=2$ and $\alpha=1/\sqrt2$, the moment relation in eq.~\eqref{eq:method-coefficient-moments} gives
\begin{equation}
 \widehat\psi_{00}=\frac{1}{2\pi},
 \qquad
 \widehat\psi_{20}=\widehat\psi_{00}\,\frac{A_{11}-1}{\sqrt2},
 \qquad
 \widehat\psi_{11}=\widehat\psi_{00}\,A_{12},
 \qquad
 \widehat\psi_{02}=\widehat\psi_{00}\,\frac{A_{22}-1}{\sqrt2},
 \label{eq:method-inlet-coeffs}
\end{equation}
All remaining coefficients are zero. The resulting inlet representation recovers $m_0=1$ and the three prescribed second moments in exact arithmetic, and the inlet stress is evaluated from this same conformation.

For the cavity flow populations, link-wise half-way bounce-back includes the moving-wall velocity correction at the lid; the side and bottom walls and the corner links use stationary bounce-back.

The coefficient transport boundaries are prescribed independently of the flow populations. Cross-slot flow uses eq.~\eqref{eq:method-inlet-coeffs} at the left and right inlets, zero normal gradients at the top and bottom outlets, and zero normal coefficient flux at solid walls; WENO stencils crossing a wall are replaced by local Lax--Friedrichs states. Poiseuille flow is periodic in the streamwise direction with homogeneous-Neumann coefficient data at the transverse walls, cavity flow uses homogeneous-Neumann coefficient data on all four walls, and four-roll flow is periodic in both directions.

For the cross-slot flow populations, initialization and pressure/non-equilibrium-extrapolation outlet reconstruction use the equilibrium without its cubic term. Collision retains the full equilibrium in eq.~\eqref{eq:method-feq}.

\subsection{Time coupling and consistency checks}
\label{sec:method-coupling}

The coupled updates exchange velocity and polymer force in two explicit sequences. Poiseuille and cross-slot flow advance the flow before the configuration, whereas four-roll and cavity flow advance the configuration first. The velocity $\boldsymbol{u}^{\mathrm{use}}$ supplies both local deformation and coefficient transport. For Poiseuille and cross-slot flow, an update starts from $f^n$, $\widehat\psi^n$, and $\boldsymbol{F}_p^n$ and follows
\begin{equation}
 \begin{aligned}
  (f^n,\boldsymbol{F}_p^n)
  &\xrightarrow{\mathrm{P\text{-}TRT}} f^{n+1},\\
  \nabla_h\boldsymbol{u}^{\mathrm{use}}
  &\longrightarrow \widehat\psi^{\mathrm{loc}}
  \longrightarrow \widehat\psi^{n+1}
  \longrightarrow \boldsymbol{A}^{n+1}
  \longrightarrow \boldsymbol{\tau}_p^{n+1}
  \longrightarrow \boldsymbol{F}_p^{n+1},
 \end{aligned}
 \label{eq:method-coupling-chain}
\end{equation}
For both cases, the flow collision uses the velocity recovered from $f^n$, with $\boldsymbol{F}_p^n$ in the half-force term. P-TRT collision, streaming, and flow-boundary treatment precede the configuration update. Poiseuille flow retains the pre-collision velocity as $\boldsymbol{u}^{\mathrm{use}}$. Cross-slot flow recovers $\boldsymbol{u}^{\mathrm{use}}$ from the resulting populations, still using $\boldsymbol{F}_p^n$ in the half-force term. Local coefficient evolution and spatial transport then provide the moments, polymer stress, and $\boldsymbol{F}_p^{n+1}$ for the next collision.

Four-roll and cavity flow first advance the local coefficients and spatial transport using the velocity retained from the preceding update. The recovered polymer force enters the current P-TRT collision together with that retained velocity. After streaming and boundary treatment, the velocity is recovered using the same new force in the half-force term and retained for the next configuration update. Neither sequence includes a within-step coupling iteration.

In all cases, the coupled updates combine implicit-Euler local evolution with sequential physical-space transport and are formally first order in time. Where advection and diffusion are advanced separately, their composition is also Lie-split. Poiseuille and cross-slot flow additionally defer the newly recovered polymer force to the next collision. The Runge--Kutta and Crank--Nicolson discretizations remain second order for their respective subproblems.

For the four-roll comparison, the reference solver advances eq.~\eqref{eq:model-oldroydb} with an independent lattice advection--diffusion discretization and periodic centred gradients. It recovers the polymer stress before flow collision and incorporates it through a second-order moment source, while the FP solver uses the stress divergence as a force. Both solvers start from the common rest state. With $T_c=L_c/U_c$, the conformation and stress after $n$ completed updates are assigned the sampling label $t_A^*=n\Delta t/T_c$ and the accompanying half-force velocity the label $t_u^*=(n+1/2)\Delta t/T_c$. The two implementations use this sampling convention to compare each field at the same labelled time and completed-update count.

\label{sec:method-diagnostics}
The diagnostics distinguish configuration normalization from fluid mass balance. The zeroth moment $m_0$ monitors local configuration probability, and recovered inlet moments check the imposed coefficient data. Population mass and volume-flux balance monitor fluid mass balance, while density variation and $\mathrm{Ma}_{\max}$ characterize departures from the incompressible limit. The cavity diagnostics also track the minimum eigenvalue of $\boldsymbol{A}$ to monitor conformation positive definiteness. These quantities accompany the field comparisons in Section~\ref{sec:results}.
\section{Results and discussion}

\label{sec:results}
We begin by assessing macroscopic recovery against analytical and independently computed reference solutions. Start-up Poiseuille flow tests the transient velocity and steady polymer stress against analytical predictions. Periodic four-roll flow extends the assessment to spatially varying extension and transport, with velocity, conformation and stress compared against an independently discretized Oldroyd--B solution. The regularized-lid cavity and cross-slot then examine the coupled response to boundary forcing: recirculation driven by a moving lid in the cavity, and the development of symmetric and asymmetric flow in an open junction.

\subsection{Start-up plane Poiseuille flow}

\label{sec:poiseuille}

Start-up plane Poiseuille flow tests the recovery of transient velocity and steady polymer stress in a controlled shear geometry. The channel is periodic in the streamwise direction, with no-slip walls separated by $H$. At $t=0$, a constant body force is applied to the fluid at rest, and the configuration density is initialized by the equilibrium Maxwellian. The reference scales are $H$ and the long-time channel-mid-plane velocity $U_c$, giving $y^*=y/H$, $u_x^*=u_x/U_c$, and $t^*=tU_c/H$; the imposed acceleration is $8\nu_0U_c/H^2$.
\begin{figure}[!htbp]
\centering
\includegraphics[width=0.60\linewidth]{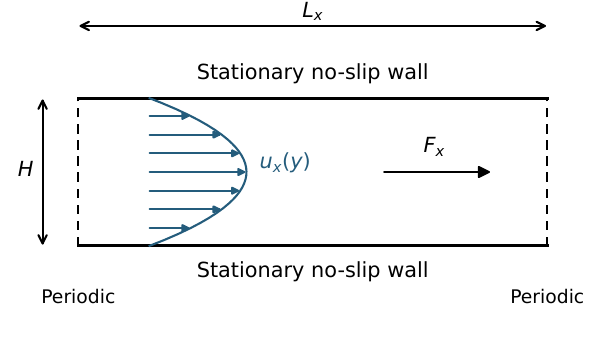}
\caption{Schematic of start-up plane Poiseuille flow. Here $L_y=H$; the
streamwise dashed boundaries are periodic and the stationary walls impose no slip.
The interior arrows illustrate the long-time velocity profile $u_x(y)$, and
$F_x$ indicates the imposed body force.}
\label{fig:poiseuille-geometry}
\end{figure}

We first compare the computed velocity with the analytical transient along the refinement path. The Waters--King solution provides this reference \cite{WatersKing1970}; in the present nondimensional variables it is
\begin{equation}
 u_x^{*,\mathrm{ana}}(y^*,t^*)
 =4y^*(1-y^*)
 -32\sum_{n=1}^{\infty}\frac{\sin(Ny^*)}{N^3}G_N(t^*),
 \qquad N=(2n-1)\pi,
 \label{eq:poiseuille-transient-velocity}
\end{equation}

where
\begin{equation}
G_N(t^*)=
\begin{cases}
\displaystyle
\exp\!\left(-\frac{\alpha_Nt^*}{2\Weiss}\right)
\left[
\cosh\!\left(\frac{\delta_Nt^*}{2\Weiss}\right)
+\frac{\gamma_N}{\delta_N}
\sinh\!\left(\frac{\delta_Nt^*}{2\Weiss}\right)
\right], & D_N>0,\\[0.9em]
\displaystyle
\exp\!\left(-\frac{\alpha_Nt^*}{2\Weiss}\right)
\left(1+\frac{\gamma_Nt^*}{2\Weiss}\right), & D_N=0,\\[0.9em]
\displaystyle
\exp\!\left(-\frac{\alpha_Nt^*}{2\Weiss}\right)
\left[
\cos\!\left(\frac{\delta_Nt^*}{2\Weiss}\right)
+\frac{\gamma_N}{\delta_N}
\sin\!\left(\frac{\delta_Nt^*}{2\Weiss}\right)
\right], & D_N<0,
\end{cases}
\label{eq:poiseuille-modal-function}
\end{equation}

and
\begin{subequations}
\label{eq:poiseuille-modal-coefficients}
\begin{align}
 \alpha_N&=1+\beta\frac{\Weiss}{\Rey}N^2,
 &D_N&=\alpha_N^2-4\frac{\Weiss}{\Rey}N^2,\\
 \delta_N&=\sqrt{|D_N|},
 &\gamma_N&=1+(\beta-2)\frac{\Weiss}{\Rey}N^2.
\end{align}
\end{subequations}

The analytical series is evaluated with $400$ modes and checked independently by a staggered finite-difference integration of the primitive velocity--stress system. For the refinement sequence below, the reference is evaluated at the achieved half-step time of each numerical profile so that the transient comparison uses the corresponding physical time.
The finite-volume sequence provides the quantitative refinement comparison. For $N_y=17$, $33$, $65$, and $129$ at $\Rey=1$, $\Weiss=1$, $\beta=0.5$, and $K=0$, the relative cell-centred profile error is
\begin{equation}
 E_2(N_y,t^*)=
 \frac{\left\|u_{\mathrm{num}}(y,t^*)-u_{\mathrm{WK}}(y,t^*)\right\|_2}
      {\left\|u_{\mathrm{WK}}(y,t^*)\right\|_2}
 \label{eq:poiseuille-error}
\end{equation}

\begin{figure}[!htbp]
\centering
\includegraphics[width=\linewidth]{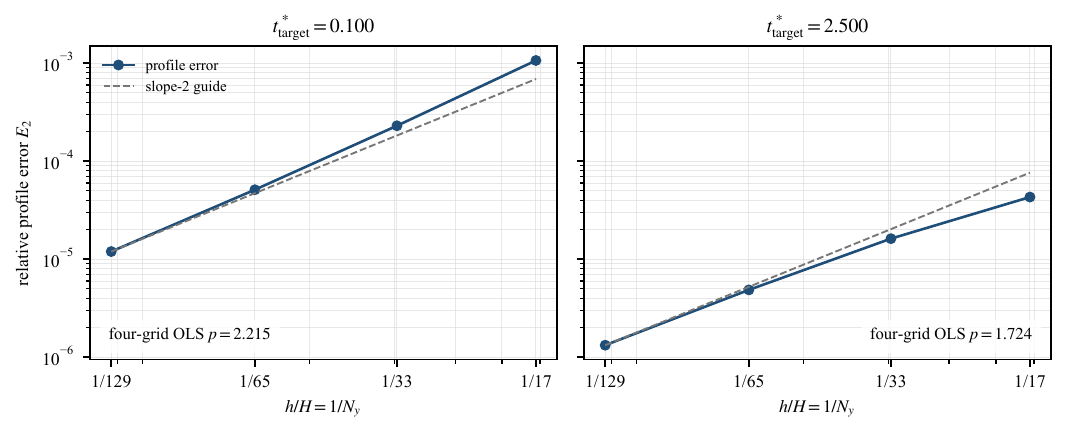}
\caption{Four-grid refinement at $Re=1$, $We=1$, $\beta=0.5$, and $K=0$:
(a) $t^*_{\mathrm{target}}=0.100$; (b) $t^*_{\mathrm{target}}=2.500$.
The four points in each panel are relative discrete $L_2$ velocity-profile
errors evaluated against Waters--King at the corresponding achieved half-step
times. Connected markers guide the eye; dashed slope-2 lines are reference guides.
The displayed OLS slopes use all four points. Spatial and temporal resolutions vary
together under diffusive scaling.}
\label{fig:poiseuille-convergence}
\end{figure}

The profile error decreases monotonically under refinement at all eight sampled times. For the two target times shown in Figure~\ref{fig:poiseuille-convergence} and Table~\ref{tab:poiseuille-convergence}, the error at $t^*=0.10$ falls from $1.07\times10^{-3}$ to $1.20\times10^{-5}$, and at $t^*=2.50$ from $4.31\times10^{-5}$ to $1.33\times10^{-6}$. Ordinary least-squares (OLS) fits over all four grids give slopes of $2.215$ and $1.724$, respectively. Along this refinement sequence, the dimensionless time step decreases in proportion to the square of the grid spacing, so these slopes describe the error decrease along this diffusive-scaling path and do not determine separate spatial or temporal orders. As a separate check of configuration-probability normalization, the largest configuration-mass departure over the $32$ sampled states is $1.04\times10^{-14}$.

\begin{table}[!htbp]
\centering
\small
\setlength{\tabcolsep}{5pt}
\renewcommand{\arraystretch}{1.12}
\caption{Velocity-profile errors and convergence slopes along the diffusive-scaling
sequence. Each reference profile is evaluated at the corresponding achieved
half-step time. Here $p_{\mathrm{pair}}$ compares consecutive grids, whereas
$p_{\mathrm{OLS}}$ is fitted over all four grids.}
\label{tab:poiseuille-convergence}
\begin{tabular}{rrrrrr}
\toprule
$t^*_{\mathrm{target}}$ & $N_y$ & $t^*_{\mathrm{achieved}}$ & $E_2$ & $p_{\mathrm{pair}}$ & $p_{\mathrm{OLS}}$ \\
\midrule
0.100 & 17 & 0.10106 & $1.07\times10^{-3}$ & -- & 2.215 \\
 & 33 & 0.10015 & $2.30\times10^{-4}$ & 2.312 &  \\
 & 65 & 0.10002 & $5.12\times10^{-5}$ & 2.217 &  \\
 & 129 & 0.10002 & $1.20\times10^{-5}$ & 2.119 &  \\
\addlinespace
2.500 & 17 & 2.50012 & $4.31\times10^{-5}$ & -- & 1.724 \\
 & 33 & 2.50015 & $1.62\times10^{-5}$ & 1.473 &  \\
 & 65 & 2.50007 & $4.87\times10^{-6}$ & 1.776 &  \\
 & 129 & 2.50001 & $1.33\times10^{-6}$ & 1.896 &  \\
\bottomrule
\end{tabular}
\par\vspace{4pt}
\begin{minipage}{0.98\linewidth}
\footnotesize\raggedright
$E_2=\|\boldsymbol{u}_{\mathrm{num}}-\boldsymbol{u}_{\mathrm{WK}}\|_2/\|\boldsymbol{u}_{\mathrm{WK}}\|_2$
on the cell-centred profile at the achieved time; $h/H=1/N_y$.
Each $p_{\mathrm{pair}}$ compares the preceding grid with the current grid.
The coupled refinement does not establish separate spatial or temporal orders.
\end{minipage}
\end{table}

\FloatBarrier

\begin{figure}[!htbp]
\centering
\includegraphics[width=\linewidth]{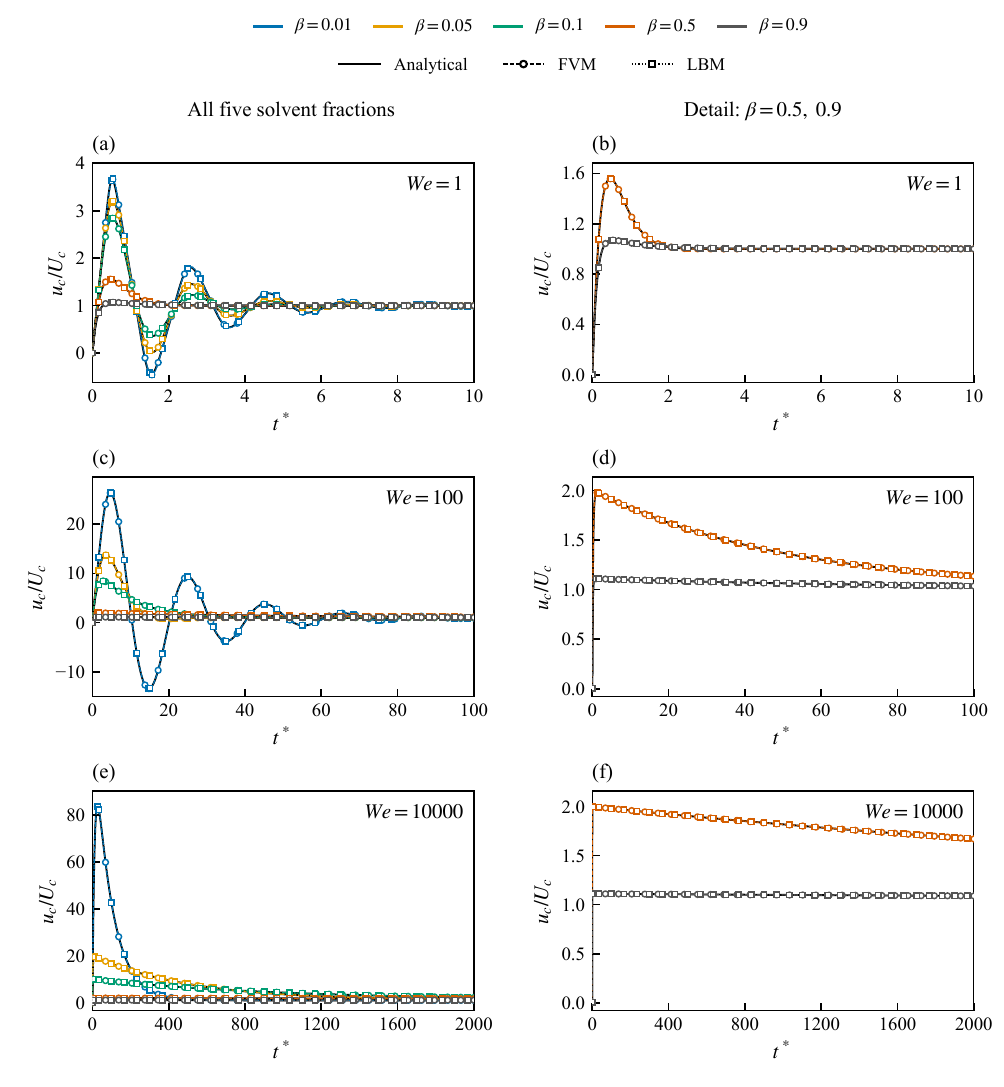}
\caption{Dimensionless velocity at the channel mid-plane versus dimensionless
time for $\beta=0.01$, $0.05$, $0.10$, $0.50$, and $0.90$.
Rows correspond to $\Weiss=1$ in (a,b), $\Weiss=100$ in (c,d), and
$\Weiss=10^4$ in (e,f), with display windows $0\leq t^*\leq10$, $100$,
and $2000$, respectively. Left panels show all five solvent fractions;
right panels show $\beta=0.50$ and $0.90$ on separate vertical scales.
Black curves are the Waters--King references. Coloured dashed lines
with open circles denote FVM; dotted lines with open squares denote regularized
scalar-LBM coefficient transport. Numerical velocities are plotted at their
achieved half-step times.}
\label{fig:poiseuille-transient-beta}
\end{figure}

The centre-velocity histories in Figure~\ref{fig:poiseuille-transient-beta} vary elasticity and solvent fraction at $\Rey=1$, $M=2$, $\alpha=\sqrt{1/2}$, and $\Lambda_s=3/16$, with $K=0$. The flow is uniform in the streamwise direction, so one periodic lattice column is sufficient; the wall separation is $L_y=33$ lattice units. The solvent flow is advanced by P-TRT LBM with half-way bounce-back. FVM and LBM in the following history and profile comparisons denote the WENO-Z/CN--ADI finite-volume and regularized scalar-lattice discretizations of the Hermite coefficients in physical space, respectively; the configuration-space update, moment recovery, and stress feedback are otherwise shared.

The reference lattice speeds are chosen to keep the computed histories at low Mach number. For the ordered solvent fractions $\beta=(0.01,0.05,0.10,0.50,0.90)$, the reference lattice velocities are $U_c=0.01$ at $\Weiss=1$, $U_c=(0.0015,0.003,0.005,0.02,0.04)$ at $\Weiss=100$, and $U_c=(0.0005,0.002,0.004,0.02,0.04)$ at $\Weiss=10^4$. The viscosities, relaxation time and body force are scaled with the reference speed while keeping $\Rey$, $\Weiss$ and $\beta$ fixed. The largest Mach number measured over all lattice nodes and time steps of the thirty histories is $0.0770$. At $\Weiss=10^4$ and $\beta=0.01$, halving $U_c$ from $5\times10^{-4}$ to $2.5\times10^{-4}$ changes the first-peak value of $u_x^*$ by less than $2.5\times10^{-6}$ for either discretization, in a separate check through $t^*=40$.

Figure~\ref{fig:poiseuille-transient-beta} compares these histories with the Waters--King solution over $0\leq t^*\leq10$, $0\leq t^*\leq100$, and $0\leq t^*\leq2000$ for $\Weiss=1$, $100$, and $10^4$, respectively. Decreasing $\beta$ strengthens the elastic transient, while increasing $\Weiss$ lengthens the relaxation time scale. Both physical-space coefficient-transport discretizations follow the reference histories, including overshoot and damped oscillations where present. Amplitude and phase differences remain at $\beta=0.01$; for scalar-LBM at $\Weiss=10^4$, the centre velocity at the end of the displayed window is $1.41\%$ above the analytical value.

\FloatBarrier

\begin{figure}[!htbp]
\centering
\includegraphics[width=\linewidth]{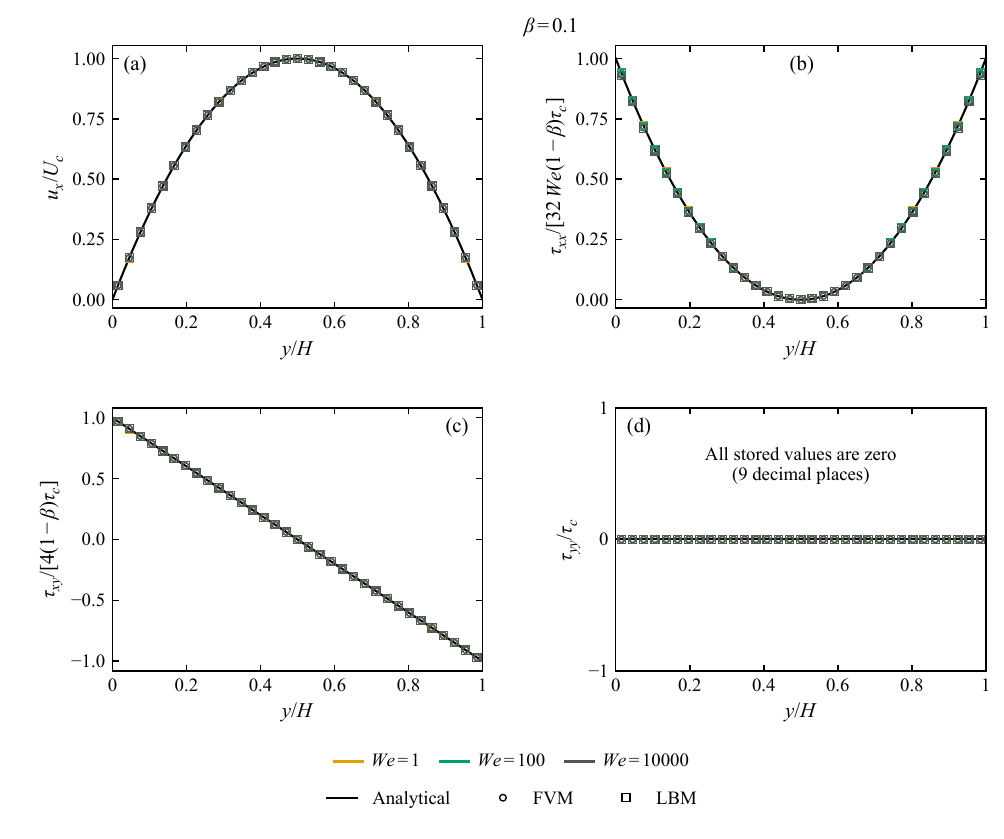}
\caption{Cross-channel velocity and polymer-stress profiles at $x=L_x/2$ for
$\beta=0.1$ and $\Weiss=1$, $100$, and $10^4$: (a) velocity;
(b) $\tau_{p,xx}^*$; (c) $\tau_{p,xy}^*$; (d) $\tau_{p,yy}^*$.
The velocity and the nonzero stresses are normalized by the maximum absolute
value of their respective analytical profiles. The $yy$ component uses the
nondimensional stress scale directly; its values are zero at nine
decimal places. In the plotted labels, $\tau_{ij}$ denotes the polymer stress
and $\tau_c=\nu_0U_c/H$. Colours identify $\Weiss$; solid lines are analytical,
open circles denote FVM, and open squares denote regularized scalar-LBM
coefficient transport. Coincident series retain their original cell-centre
positions. FVM and LBM identify only the physical-space discretization of the
Fokker--Planck coefficients; the solvent flow is advanced by P-TRT LBM in both cases.}
\label{fig:poiseuille-validation}
\end{figure}

\begin{figure}[!htbp]
\centering
\includegraphics[width=\linewidth]{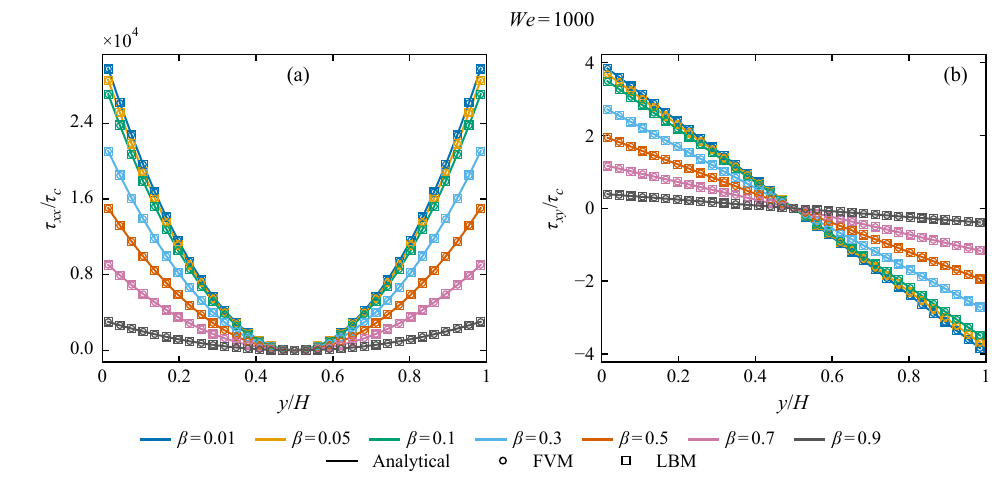}
\caption{Unnormalised polymer-stress profiles at $\Weiss=1000$ for
$\beta=0.01$, $0.05$, $0.10$, $0.30$, $0.50$, $0.70$, and $0.90$:
(a) $\tau_{p,xx}^*$; (b) $\tau_{p,xy}^*$. Colours identify $\beta$.
Solid lines are analytical; open circles and open squares denote FVM and
regularized scalar-LBM coefficient transport, respectively. The plotted
columns retain their nondimensional amplitudes, with $\tau_c=\nu_0U_c/H$.}
\label{fig:poiseuille-stress-beta}
\end{figure}

\begin{figure}[!htbp]
\centering
\includegraphics[width=\linewidth]{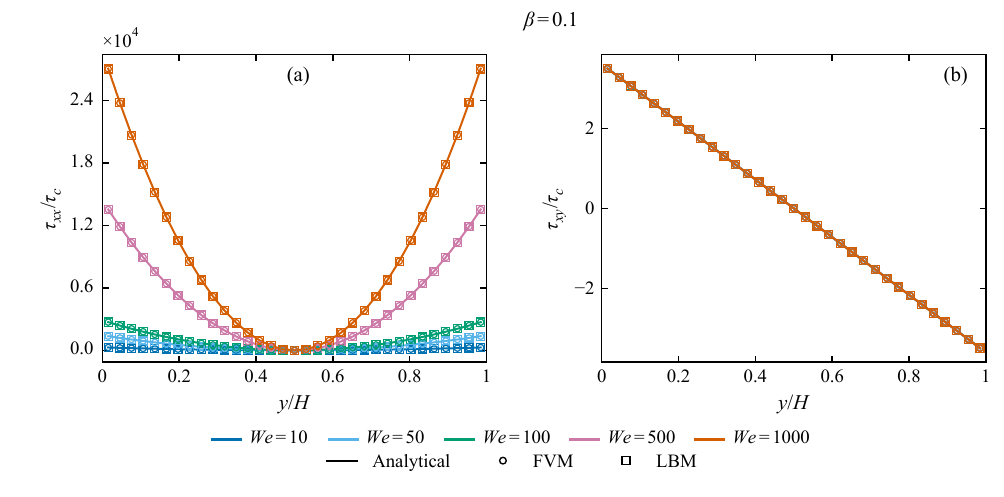}
\caption{Unnormalised polymer-stress profiles at $\beta=0.1$ for
$\Weiss=10$, $50$, $100$, $500$, and $1000$: (a) $\tau_{p,xx}^*$;
(b) $\tau_{p,xy}^*$. Colours identify $\Weiss$. Solid lines are analytical;
open circles and open squares denote FVM and regularized scalar-LBM
coefficient transport, respectively. The analytical $xy$ profiles coincide at
this fixed $\beta$. The plotted stress
scale is $\tau_c=\nu_0U_c/H$.}
\label{fig:poiseuille-stress-we}
\end{figure}

The transient comparison assesses the resulting flow, while the long-time stress profiles directly assess the recovered stress feedback. The profile comparisons in Figures~\ref{fig:poiseuille-validation}--\ref{fig:poiseuille-stress-we} use $L_y=33$ and $U_c=0.05$ in lattice units. The long-time velocity is $u_x^{*,\mathrm{ana}}=4y^*(1-y^*)$. With $\nu_0U_c/H$ as the stress scale, the corresponding Hookean polymer stresses are
\begin{equation}
 \tau_{p,xx}^{*,\mathrm{ana}}=2\Weiss(1-\beta)
 \left(\frac{\partial u_x^*}{\partial y^*}\right)^2,
 \qquad
 \tau_{p,xy}^{*,\mathrm{ana}}=(1-\beta)
 \frac{\partial u_x^*}{\partial y^*},
 \qquad
 \tau_{p,yy}^{*,\mathrm{ana}}=0.
 \label{eq:poiseuille-steady-stress}
\end{equation}

The steady profiles recover the cross-channel structure predicted by the analytical solution: a parabolic velocity, antisymmetric shear stress $\tau_{p,xy}^*$, and symmetric normal stress $\tau_{p,xx}^*$ with its minimum at the mid-plane. Figure~\ref{fig:poiseuille-validation} compares the two physical-space discretizations at $\beta=0.1$ and $\Weiss=1$, $100$, and $10^4$. For each nonzero field, both numerical and analytical profiles use the maximum absolute value of the analytical profile as the normalization scale, retaining relative amplitude deviations across $\Weiss$. The $\tau_{p,yy}^*$ values are zero at the saved precision of nine decimal places, consistent with the analytical zero reference.

The unnormalised profiles distinguish the effects of solvent fraction and elasticity on stress amplitude. Equation~\eqref{eq:poiseuille-steady-stress} predicts that both $\tau_{p,xx}^*$ and $\tau_{p,xy}^*$ scale with $1-\beta$, whereas only the normal-stress amplitude scales with $\Weiss$ for the same parabolic velocity profile. At fixed $\Weiss=1000$, the FVM and scalar-LBM profiles follow the decrease in both stress amplitudes as $\beta$ increases (Figure~\ref{fig:poiseuille-stress-beta}). At fixed $\beta=0.1$, the profiles for $\Weiss=10$, $50$, $100$, $500$, and $1000$ show increasing normal stress and a shear-stress profile independent of $\Weiss$ (Figure~\ref{fig:poiseuille-stress-we}).

\FloatBarrier

\FloatBarrier

The four-roll case extends the comparison to spatially varying extension and transport with two-way stress feedback.
\FloatBarrier

\subsection{Four-roll mill flow}
\label{sec:fourroll}

\begin{figure}[!htbp]
\centering
\includegraphics[width=0.58\linewidth]{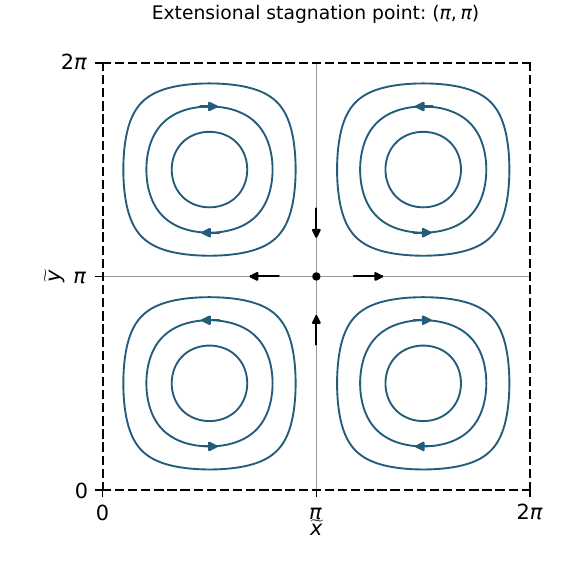}
\caption{Four-roll topology of the Newtonian velocity field in eq.~\eqref{eq:fourroll-newtonian-velocity}. Curves are contours of $\psi=\sin\widetilde{x}\sin\widetilde{y}$, arrows show the flow direction, and the central dot marks the extensional stagnation point $(\pi,\pi)$. Dashed outer boundaries are periodic.}
\label{fig:fourroll-geometry}
\end{figure}

Periodic four-roll flow tests whether the configuration-based calculation
recovers the macroscopic response when polymer stretching, spatial transport,
and two-way stress feedback vary across the domain.
For the Hookean model, the FP equation and the direct Oldroyd--B formulation
share the same macroscopic target.  The separately implemented Oldroyd--B
calculation is therefore used as an independent macroscopic comparator for the
FP response to spatially varying extension around the central stagnation point.
The doubly periodic domain is $\Omega=[0,2\pi)\times[0,2\pi)$.  On an
$N\times N$ cell-centred grid,
\begin{equation}
 \widetilde{x}_i=\left(i-\tfrac12\right)\frac{2\pi}{N},
 \qquad
 \widetilde{y}_j=\left(j-\tfrac12\right)\frac{2\pi}{N},
 \qquad i,j=1,\ldots,N.
 \label{eq:fourroll-grid}
\end{equation}
Mapping the $2\pi$ box to $N$ lattice cells gives the reference scales
$L_c=N/(2\pi)$ and $U_c=U_0$ in lattice units.  The input Mach number fixes
$U_0=0.1c_s$.
The four-roll topology is generated by the equivalent body force
\begin{equation}
 \boldsymbol{F}_e
 =2F\left(\sin\widetilde{x}\cos\widetilde{y},
          -\cos\widetilde{x}\sin\widetilde{y}\right),
 \qquad
 F=U_0\nu_s\left(\frac{2\pi}{N}\right)^2 .
 \label{eq:fourroll-force}
\end{equation}
In the Newtonian limit the corresponding velocity topology is
\begin{equation}
 \boldsymbol{u}_{\mathrm N}
 =U_0\left(\sin\widetilde{x}\cos\widetilde{y},
          -\cos\widetilde{x}\sin\widetilde{y}\right).
 \label{eq:fourroll-newtonian-velocity}
\end{equation}
Figure~\ref{fig:fourroll-geometry} shows the streamlines of this Newtonian velocity field and the central extensional stagnation point.

\FloatBarrier

The Fokker--Planck and direct Oldroyd--B calculations start from the common rest state $\rho=1$,
$\boldsymbol u=0$, $\boldsymbol A=\boldsymbol I$, and
$\boldsymbol\tau_p=0$; the FP state represents $\boldsymbol A=\boldsymbol I$
through the equilibrium Gaussian coefficients.  Both calculations use
$\Rey=1$, $\beta=2/3$, and $\Sc=10^5$, with input Mach number $0.1$; the FP representation uses
$M=2$.  We first vary elasticity on the finest grid, $N=257$, over
$\Weiss\in\{0.1,0.3,0.6,1.0\}$, and compare the paired fields near
$t^*=10$ ($t_A^*=10.00060$ for conformation and stress,
$t_u^*=10.00130$ for the half-force velocity).  A separate low-elasticity
sequence then varies the grid to identify how the discrepancy changes with
resolution.
For each FP/Oldroyd--B comparison, the relative field difference for
$X\in\{\boldsymbol u,\boldsymbol A,\boldsymbol\tau_p\}$ is
\begin{equation}
 E_X=
 \frac{\left\|X_{\mathrm{FP}}-X_{\mathrm{OB}}\right\|_{2,\Omega}}
      {\left\|X_{\mathrm{OB}}\right\|_{2,\Omega}},
 \label{eq:fourroll-macro-error}
\end{equation}
with Euclidean or Frobenius component norms as appropriate. Since $E_A$ and $E_{\tau}$ use different reference norms in their denominators, a larger $E_{\tau}$ need not represent an additional stress discrepancy.
These norms measure full-domain differences between two discretizations of
the common macroscopic target, rather than errors relative to an exact solution.

For the four comparisons at $N=257$ and near $t^*=10$, spanning
$\Weiss\in\{0.1,0.3,0.6,1.0\}$, the largest relative velocity difference
in Table~\ref{tab:fourroll-finest-errors} is $3.09\times10^{-4}$.
The conformation and polymeric-stress differences increase with $\Weiss$,
reaching $2.82\times10^{-2}$ and $3.84\times10^{-2}$, respectively,
at $\Weiss=1$.
At this largest tested elasticity, a small velocity difference thus coexists
with larger relative differences in the constitutive fields.
The velocity and constitutive fields therefore need to be assessed separately.

\begin{table}[!htbp]
\centering
\caption{Relative differences between the FP and direct Oldroyd--B fields for
the finest-grid comparison. The full-domain relative norms are defined in
eq.~\eqref{eq:fourroll-macro-error}.}
\label{tab:fourroll-finest-errors}
\begin{tabular}{@{}cccc@{}}
\toprule
$\Weiss$ & $E_u$ & $E_A$ & $E_{\tau}$ \\
\midrule
$0.1$ & $1.25\times10^{-4}$ & $8.78\times10^{-5}$ & $1.31\times10^{-3}$ \\
$0.3$ & $1.95\times10^{-4}$ & $2.90\times10^{-4}$ & $1.40\times10^{-3}$ \\
$0.6$ & $2.85\times10^{-4}$ & $1.41\times10^{-3}$ & $3.22\times10^{-3}$ \\
$1.0$ & $3.09\times10^{-4}$ & $2.82\times10^{-2}$ & $3.84\times10^{-2}$ \\
\bottomrule
\end{tabular}
\end{table}

\begin{figure}[!htbp]
\centering
\includegraphics[width=170mm]{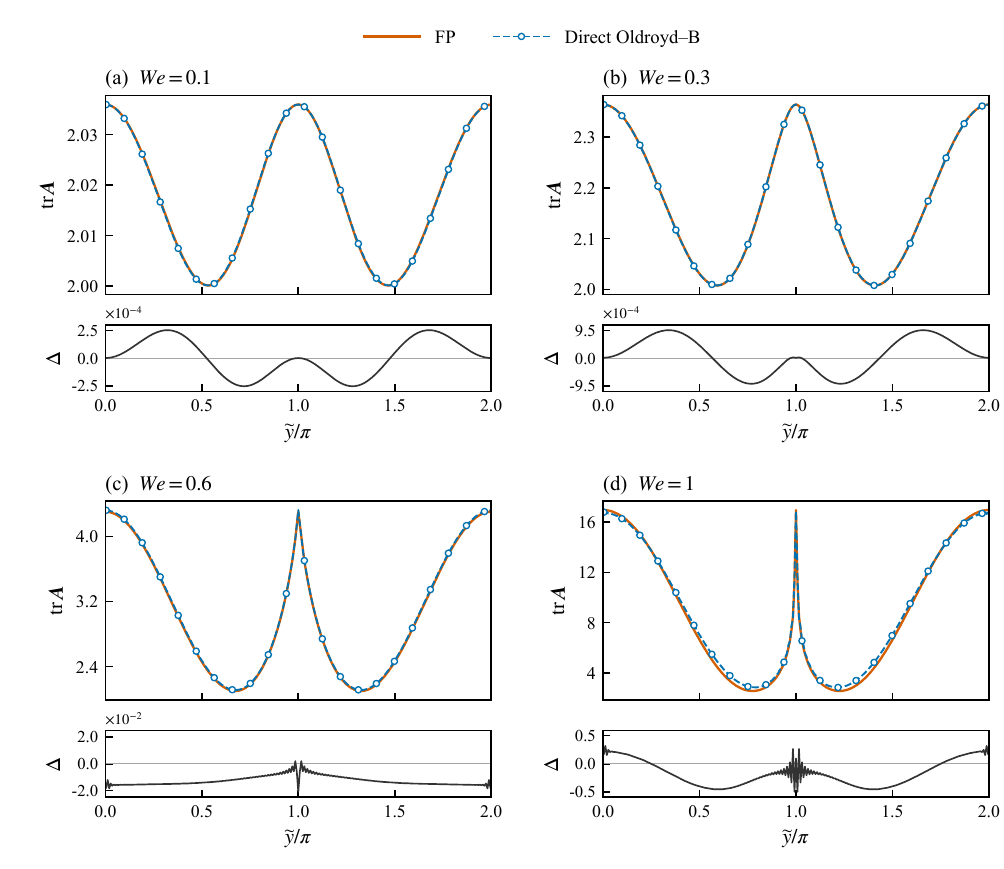}
\caption{Centreline conformation trace at $\widetilde{x}/\pi=1$ for the FP and
separately implemented direct Oldroyd--B calculations on the $N=257$ grid.
The four panels show $\Weiss=0.1$, $0.3$, $0.6$, and $1.0$ at
$t_A^*=10.00060$; the ordinate is the dimensionless
$\operatorname{tr}\boldsymbol A$ and the position is $\widetilde{y}/\pi$.
Solid orange lines denote FP; dashed blue lines with open circles denote
direct Oldroyd--B. The curves are plotted at the common output locations;
markers identify selected sampling locations. The lower strips show
$\Delta=\operatorname{tr}\boldsymbol A_{\mathrm{FP}}-
\operatorname{tr}\boldsymbol A_{\mathrm{OB}}$, with a separate scale in each
strip. Narrow peaks and local oscillations are visible.}
\label{fig:fourroll-profiles}
\end{figure}

The conformation-trace profiles nearly coincide at lower $\Weiss$, while differences become visible near and away from the narrow central peak at higher $\Weiss$ (Figure~\ref{fig:fourroll-profiles}). Both calculations retain this peak, but their signed trace differences show local oscillations around it. At $\Weiss=1$, the profiles separate over a broader region, with negative differences away from the peak and positive differences near the periodic ends.

\begin{figure}[!htbp]
\centering
\includegraphics[width=170mm]{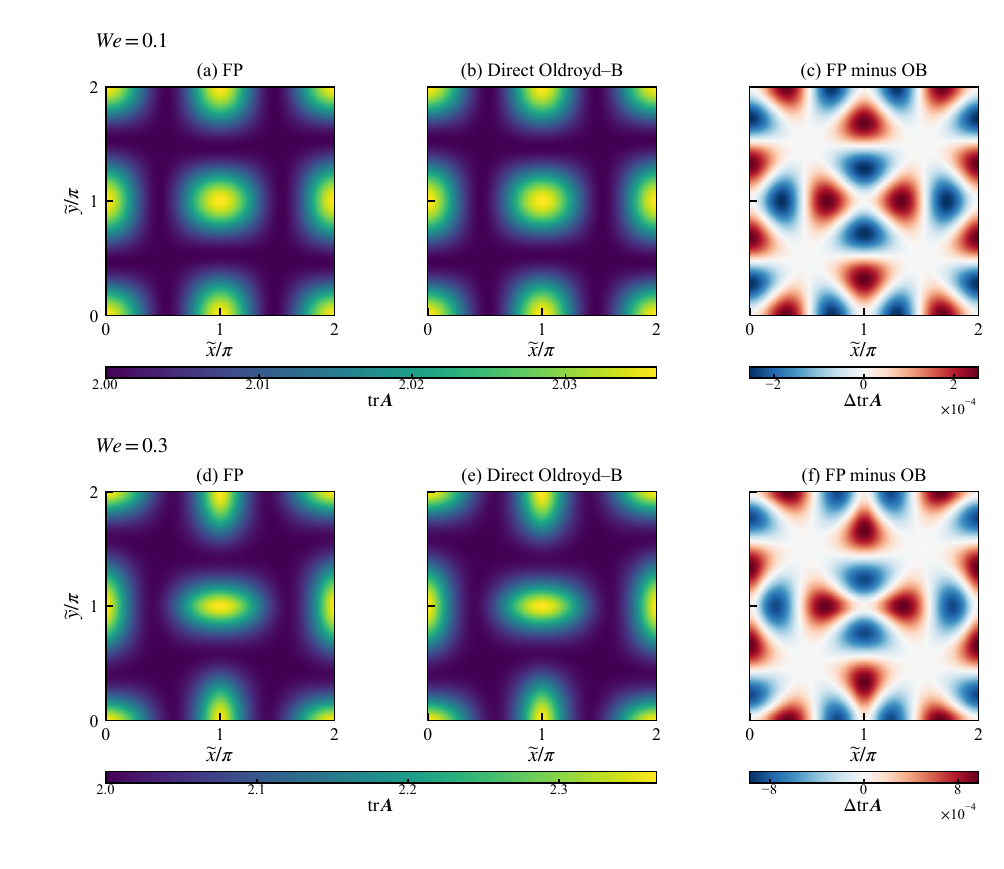}
\caption{Full-domain conformation trace and signed FP/direct-Oldroyd--B
trace difference for $\Weiss=0.1$ and $0.3$ at $N=257$ and
$t_A^*=10.00060$. Coordinates are $\widetilde{x}/\pi$ and
$\widetilde{y}/\pi$. Within each row, FP and direct Oldroyd--B use a shared
colour scale, and the FP-minus-OB difference uses a separate symmetric scale
centred at zero. The two rows have different scales.
Each cell shows the cell-centred value; no interpolation is applied.}
\label{fig:fourroll-fields-low}
\end{figure}

\begin{figure}[!htbp]
\centering
\includegraphics[width=170mm]{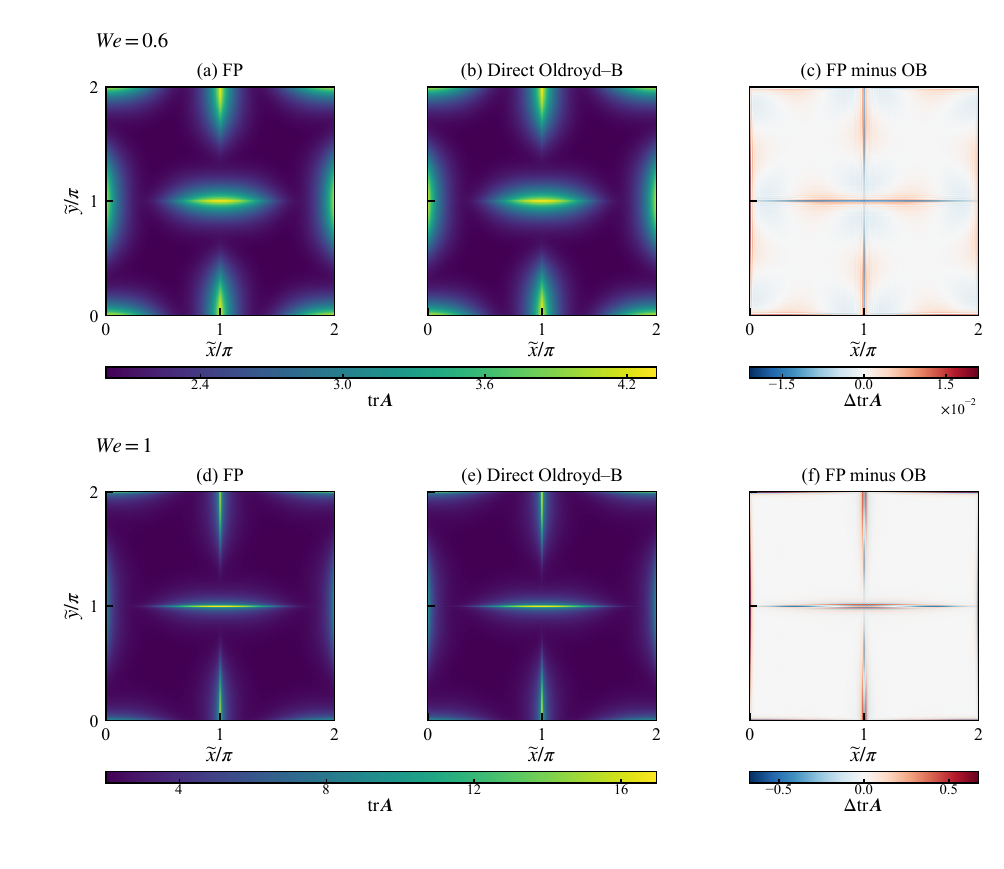}
\caption{Full-domain conformation trace and signed FP/direct-Oldroyd--B
trace difference for $\Weiss=0.6$ and $1.0$ at $N=257$ and
$t_A^*=10.00060$. Coordinates are $\widetilde{x}/\pi$ and
$\widetilde{y}/\pi$ in the periodic square. Within each row, the FP and direct
Oldroyd--B panels share one colour scale; the FP-minus-OB difference has a
separate symmetric scale centred at zero. Colour scales differ between rows.
Each coloured cell represents its cell-centred value without interpolation.}
\label{fig:fourroll-fields-high}
\end{figure}

The full-domain maps locate trace differences outside the selected centreline. At $\Weiss=0.1$ and $0.3$, the FP and direct Oldroyd--B distributions remain close, although the signed maps reveal positive and negative differences across the domain (Figure~\ref{fig:fourroll-fields-low}). At $\Weiss=0.6$ and $1.0$, both calculations exhibit narrow high-trace regions at the central structure and near the periodic boundaries, with local differences in their magnitudes (Figure~\ref{fig:fourroll-fields-high}). Full-tensor and stress differences are quantified separately by the field norms.

\FloatBarrier

At $\Weiss=0.1$, a separate sequence on $N=65$, 129, and 257 grids examines
how the FP/direct-Oldroyd--B agreement changes with resolution.
The velocity, conformation, and stress differences all decrease across the
three grids (Table~\ref{tab:fourroll-grid-trend}).
Between $N=129$ and 257, the adjacent-grid slopes of these differences are
1.781 for velocity and approximately 0.669 for both conformation and stress.
The two implementations thus approach each other under refinement at this
low-elasticity condition; the slopes of their differences do not determine
the convergence order of either implementation separately.
\begin{table}[!htbp]
\centering
\caption{Three-grid FP/direct-Oldroyd--B discrepancy trend at the low-elasticity
condition. The two $p$ columns are adjacent-grid slopes; the paired
implementations share each comparison clock.}
\label{tab:fourroll-grid-trend}
\begin{tabular}{@{}cccccc@{}}
\toprule
Metric & $N=65$ & $N=129$ & $N=257$ & $p_{65,129}$ & $p_{129,257}$ \\
\midrule
$E_u$ & $1.54\times10^{-3}$ & $4.26\times10^{-4}$ & $1.25\times10^{-4}$ & $1.881$ & $1.781$ \\
$E_A$ & $2.43\times10^{-4}$ & $1.39\times10^{-4}$ & $8.78\times10^{-5}$ & $0.813$ & $0.669$ \\
$E_{\tau}$ & $3.64\times10^{-3}$ & $2.08\times10^{-3}$ & $1.31\times10^{-3}$ & $0.815$ & $0.669$ \\
\bottomrule
\end{tabular}
\end{table}

\begin{figure}[!htbp]
\centering
\includegraphics[width=170mm]{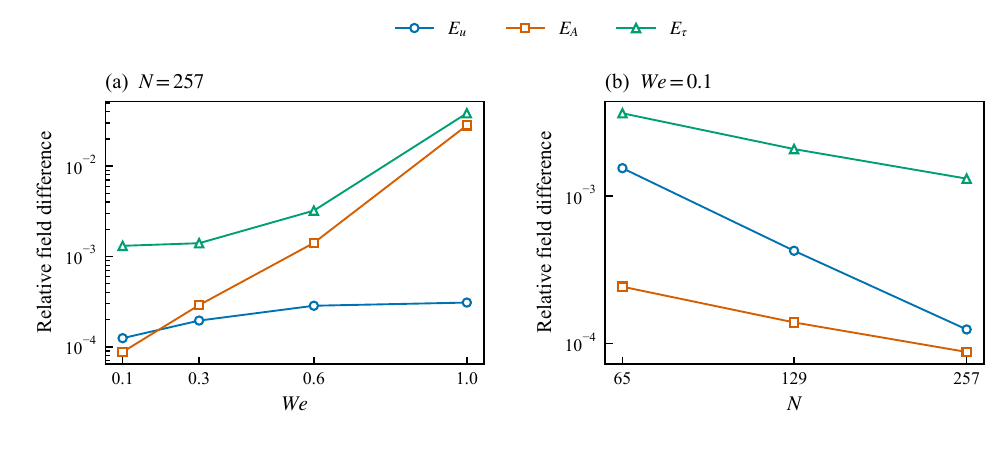}
\caption{Relative differences between the discrete FP and direct
Oldroyd--B fields. (a) The four $\Weiss$ values at $N=257$;
(b) the three-grid trend at $\Weiss=0.1$. The metrics are $E_u$, $E_A$, and
$E_{\tau}$ from eq.~\eqref{eq:fourroll-macro-error}, with direct Oldroyd--B
field norms in the denominators. Conformation/stress and half-force velocity
are paired near $t^*=10$, with matching FP/OB sampling times for each
variable on each grid and slight time differences between grids.
Connecting lines join sampled conditions without fitting a law.
Panel (b) shows the inter-implementation trend at a single low-$\Weiss$
condition.}
\label{fig:fourroll-trends}
\end{figure}

Figure~\ref{fig:fourroll-trends} summarizes the elasticity and grid trends from Tables~\ref{tab:fourroll-finest-errors} and \ref{tab:fourroll-grid-trend}.

\FloatBarrier

The diagnostics assess normalization and internal consistency in the sampled
states.  The FP configuration probability mass, evaluated by quadrature,
deviates from unity by at most $1.29\times10^{-12}$.
Across the sampled FP and direct Oldroyd--B states, the smallest eigenvalue
of $\boldsymbol A$ is $0.383$, supporting positive-definite conformation.
The maximum reflection residual, $9.55\times10^{-13}$, indicates that the
sampled fields preserve the four-roll reflection symmetries.
The maximum relative Kramers-map residual, $2.23\times10^{-16}$, checks the
internal algebraic consistency of the moment--stress relation in the two
calculations.  The maximum density deviation, $2.22\times10^{-3}$, and maximum
Mach number, $0.0666$, are consistent with the low-Mach, nearly incompressible
flow description.

Field norms and spatial comparisons quantify FP/direct-Oldroyd--B
macroscopic agreement and remaining differences for the tested grids,
parameters, and sampling times at $\Weiss\leq1$.
The lid-driven cavity next examines coupling between configuration and flow
with a moving wall and closed recirculation.

\FloatBarrier

\subsection{Lid-driven cavity flow}
\label{sec:cavity}

The lid-driven cavity illustrates the coupled flow and configuration response
under solid-wall confinement. Its moving lid and stationary walls generate
closed recirculation and near-wall velocity gradients, extending the periodic
four-roll setting to a flow driven through its boundary.
The square cavity occupies $\Omega=[0,L_c]\times[0,L_c]$. The side and bottom
walls are stationary, whereas the top wall follows the regularized R1 lid
profile used by Comminal et al.~\cite{ComminalSpangenbergHattel2015}; the two
top corners remain stationary. With $s=x/L_c$ and $t^*=tU_0/L_c$, the imposed
velocity is

\begin{equation}
 u_x(s,t)=8U_0\left[1+\tanh\left(8\left(t^*-\tfrac12\right)\right)\right]
 s^2(1-s)^2,
 \qquad u_y=0 .
 \label{eq:cavity-lid-velocity}
\end{equation}

The regularization makes the lid speed vanish at both corners and approach a
centreline maximum of $U_0$ after the start-up interval.  We use one
$N=65$ grid to examine the coupled response at $\Rey=1$, $\beta=0.5$,
$\Ma=0.1$, $\Weiss=0.5$, and $\Sc=10^5$; the retained configuration basis
has $M=2$ and $\alpha=1/\sqrt{2}$, and the solvent relaxation parameter is
$\Lambda_s=3/16$.  The wall and coefficient-diffusion treatments follow
Section~\ref{sec:method-boundaries}.

The terminal-state diagnostic compares consecutive samples indexed by $n$,
separated by $\lfloor T_c/\Delta t\rfloor\Delta t$, approximately one
convective time $T_c=L_c/U_0$ (1125 flow steps here):

\begin{equation}
 E_u^n=\frac{\lVert\boldsymbol{u}^{n}-\boldsymbol{u}^{n-1}\rVert_\infty}{U_0},
 \qquad
 E_\tau^n=\lVert\boldsymbol{\tau}_p^{n}-\boldsymbol{\tau}_p^{n-1}\rVert_\infty,
 \qquad
 E^n=\max\!\left(E_u^n,E_\tau^n\right).
 \label{eq:cavity-steady-error}
\end{equation}

Here $\boldsymbol{\tau}_p$ is the kinematic polymer stress in lattice units,
used without further normalization in $E_\tau^n$.
The terminal criterion is $t^*\geq20$ and $E^n<5\times10^{-5}$.
The sampled state at $t^*=20.98$ satisfies this criterion and supplies both the
velocity and conformation views below.
At this state, $E^n=2.84\times10^{-9}$, the maximum configuration-space
probability-mass deviation is $2.44\times10^{-13}$, and the minimum
eigenvalue of $\boldsymbol A$ is $0.318$.

\begin{figure}[!htbp]
\centering
\includegraphics[width=170mm]{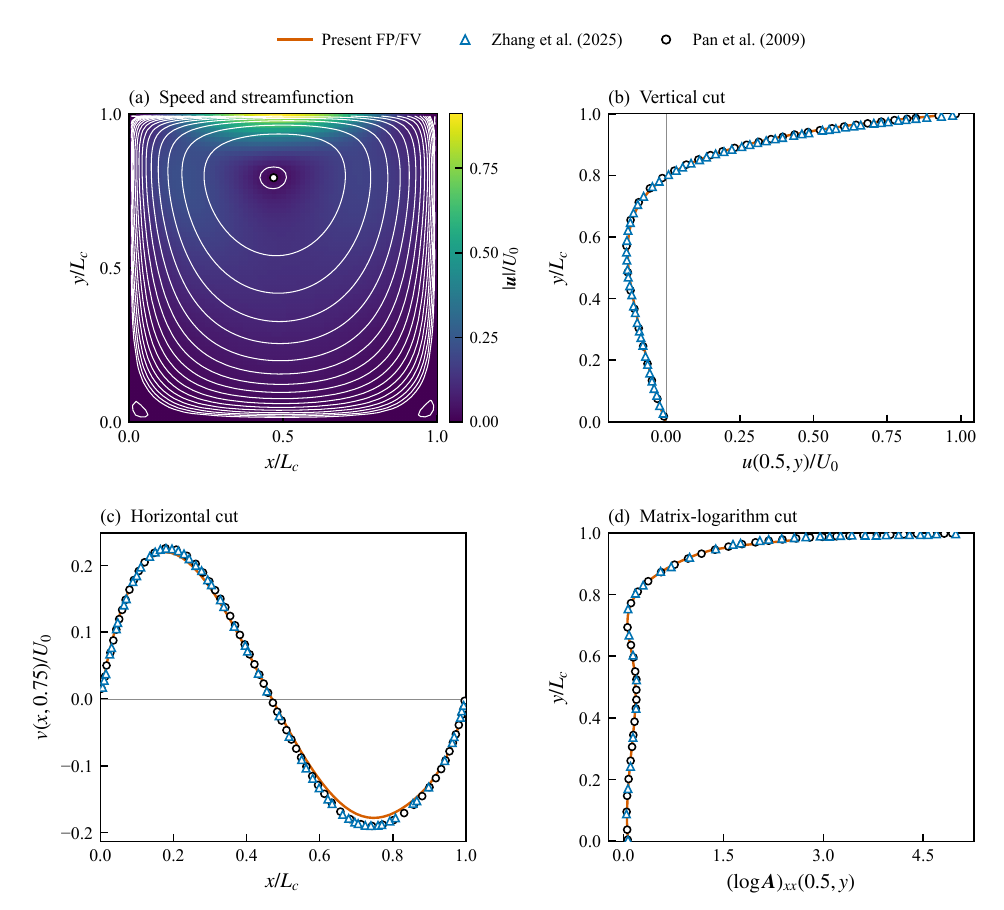}
\caption{Lid-driven cavity response with a qualitative reference overlay at
$\Weiss=0.5$ and $\beta=0.5$ under the R1 lid condition.
The present calculation uses $N=65$, $\Rey=1$,
$\Sc=10^5$, and $t^*=20.98$.
Coordinates and velocities are normalized by $L_c$
and $U_0$, respectively. (a) Speed magnitude and contours of the computed
discrete streamfunction; the white circle identifies its internal minimum
at $(x/L_c,y/L_c)=(0.469,0.792)$.
(b) $u(0.5,y)/U_0$; (c) $v(x,0.75)/U_0$;
(d) the matrix-logarithm component $(\log\boldsymbol A)_{xx}(0.5,y)$.
Solid lines show the 65 cell-centred samples.  Open symbols show the
author-supplied digitized reference profiles identified in the text. The $y/L_c=0.75$ cut is
interpolated at cell centres, with wall endpoints omitted.}
\label{fig:cavity-response}
\end{figure}

The velocity and conformation profiles in Figure~\ref{fig:cavity-response}
are compared with author-supplied digitized profiles attributed to
Zhang et al.~\cite{ZhangShuWangLiu2025} and
Pan et al.~\cite{PanHaoGlowinski2009}, included for qualitative comparison.

\begin{figure}[!htbp]
\centering
\includegraphics[width=170mm]{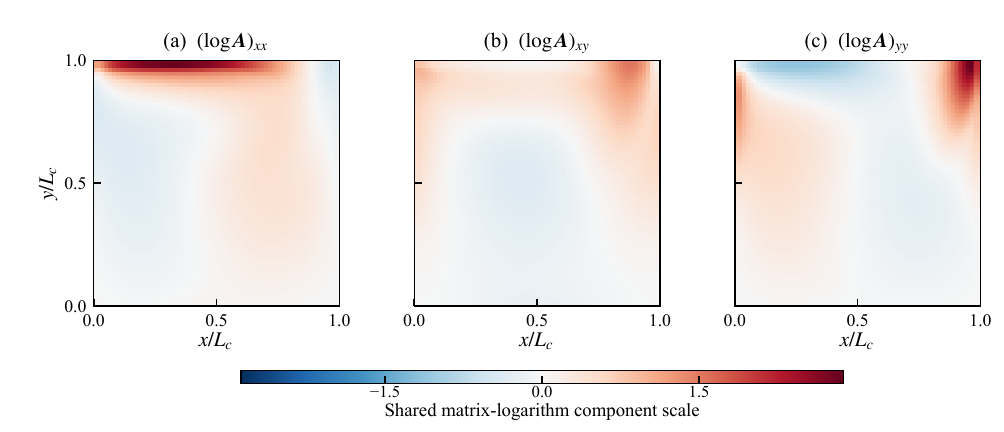}
\caption{Matrix-logarithm components (a) $(\log\boldsymbol A)_{xx}$,
(b) $(\log\boldsymbol A)_{xy}$, and (c) $(\log\boldsymbol A)_{yy}$ for the
same R1 lid-driven cavity state as Figure~\ref{fig:cavity-response}:
$N=65$, $\Rey=1$, $\Weiss=0.5$, $\beta=0.5$, $\Sc=10^5$, and
$t^*=20.98$. The axes are $x/L_c$ and $y/L_c$, and the tensor components
are dimensionless. Cell-centred values use one symmetric colour range covering
the extrema of all three components.}
\label{fig:cavity-log-fields}
\end{figure}

The flow develops one dominant recirculation cell, with the minimum of the computed discrete streamfunction at $(0.469,0.792)$. The vertical profile $u(0.5,y)$ passes from negative return flow to positive motion near the lid. Across the cell, $v(x,0.75)$ changes sign, with a smaller magnitude of the negative-velocity minimum than in the reference profiles. Along the vertical cut, the matrix-logarithm component $(\log\boldsymbol A)_{xx}(0.5,y)$ rises sharply near the lid.
The strongest configuration response is concentrated near the upper boundary
(Figure~\ref{fig:cavity-log-fields}). The positive
$(\log\boldsymbol A)_{xx}$ band extends beneath the lid, while the strongest
positive $(\log\boldsymbol A)_{yy}$ values lie near the upper-right wall.
The $(\log\boldsymbol A)_{xy}$ component has a smaller magnitude and changes
sign between the cavity interior and the upper region.
These components are obtained from the symmetric matrix logarithm:
if $\boldsymbol A=\boldsymbol Q\operatorname{diag}(\lambda_1,\lambda_2)
\boldsymbol Q^{\mathsf T}$, then
$\log\boldsymbol A=\boldsymbol Q\operatorname{diag}(\log\lambda_1,
\log\lambda_2)\boldsymbol Q^{\mathsf T}$.

Together, the profiles and full-domain fields show how the boundary-driven
recirculation is accompanied by a spatially nonuniform conformation response.
\FloatBarrier

\subsection{Cross-slot flow}
\label{sec:crossslot}

The cross-slot examines symmetric and asymmetric responses of the coupled
calculation in an open geometry.
Streams entering from opposite horizontal arms meet near the central stagnation
region and leave through the vertical arms, bringing open boundaries, solid
walls, and localized polymer stretching into the same flow.
We separate three comparisons: elasticity at $\Weiss=0.55$ and $0.60$,
initial-seed sign at $\Weiss=0.75$, and W41 versus W61 resolution over a
common observation window.

\begin{figure}[!htbp]
\centering
\includegraphics[width=0.75\linewidth]{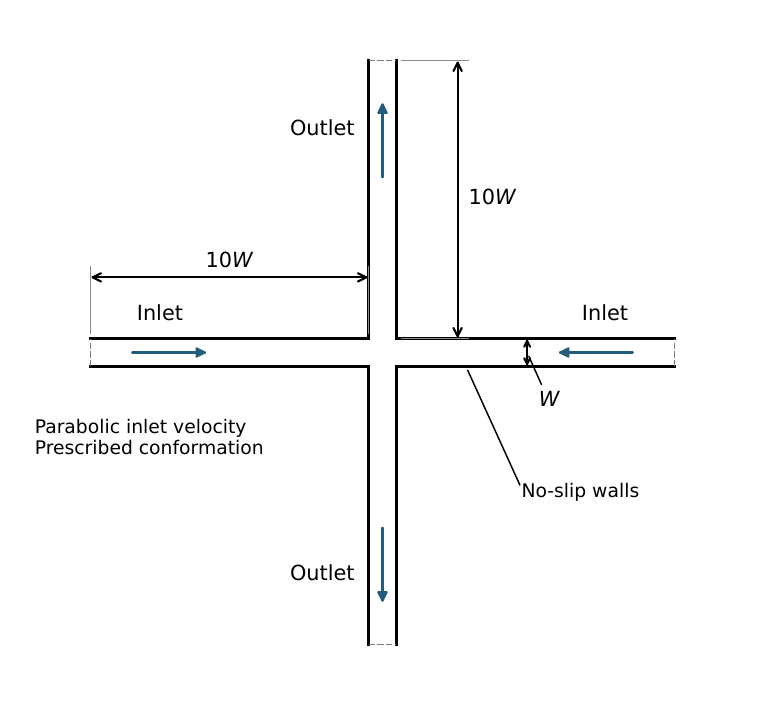}
\caption{Cross-slot geometry. The horizontal inlets prescribe a parabolic
velocity profile and conformation; the vertical arms are outlets.  From the edge of the central
$W\times W$ junction, each arm extends $10W$; its outer cross-section is $10.5W$
from the centre.}
\label{fig:crossslot-geometry}
\end{figure}

\label{sec:crossslot-setup}
The planar cross-slot is formed by two perpendicular channels of width $W$.
From the edge of the central $W\times W$ junction, each arm extends $10W$
outward, placing its outer cross-section $10.5W$ from the centre
(Figure~\ref{fig:crossslot-geometry}).
Fully developed parabolic profiles with peak speed $U_0$ are imposed at the
horizontal inlets, the vertical hydrodynamic outlets use
pressure/non-equilibrium extrapolation, and the remaining boundaries are
no-slip walls.
The prescribed inlet conformation is represented by the moment-exact coefficients
of eq.~\eqref{eq:method-inlet-coeffs}; the coefficients use zero normal gradients
at the outlets and zero normal flux at solid walls.
The corresponding boundary implementation is specified in
Section~\ref{sec:method-boundaries}.

The calculations use the groups in eq.~\eqref{eq:crossslot-groups},
with $\Weiss$ set separately for each case.

\begin{equation}
 \begin{aligned}
  \Rey&=\frac{U_0W}{\nu_0}=1,
  &\Weiss&=\frac{\lambda U_0}{W},
  &\beta&=\frac{\nu_s}{\nu_0}=\frac{1}{9},\\
  \Sc&=10^5,
  &U_0&=0.003125.
 \end{aligned}
 \label{eq:crossslot-groups}
\end{equation}

The coefficient diffusivity is $K=\nu_s/\Sc$, and time is measured as
$t^*=tU_0/W$.
The labels W41 and W61 denote grids with 41 and 61 cells, respectively,
across the channel width $W$; all five cases use $M=2$.
Table~\ref{tab:crossslot-cmin} lists their parameters and observation windows.
On W41, S1 and S2 vary $\Weiss$ from $0.55$ to $0.60$ with the same positive
seed $\epsilon=+10^{-4}$.
S3 and S4 hold $\Weiss=0.75$ and W41 fixed while changing the seed from
$-10^{-4}$ to $+10^{-4}$.
S5 retains the parameters and positive seed of S4 on W61.
The W41 simulations are evaluated through $t^*=400$, and W61 through $t^*=130$.

Two complementary observables distinguish the magnitude and direction of an
asymmetric response.
The unsigned mirror-asymmetry order parameter is the normalized residual of
the vertical velocity.
With $(\mathcal R_yu_y)(x,y)=u_y(x,-y)$ and $\Omega_f$ the fluid-node mask,

\begin{equation}
 A(t)=
 \frac{\left\|u_y(\boldsymbol{x},t)+\mathcal R_yu_y(\boldsymbol{x},t)\right\|_{2,\Omega_f}}
      {\sqrt{2}\,\left\|u_y(\boldsymbol{x},t)\right\|_{2,\Omega_f}}.
 \label{eq:crossslot-order}
\end{equation}

Because $A$ is unsigned, the flow split from the labelled left inlet supplies
the signed order parameter

\begin{equation}
 D_Q=\frac{Q_1-Q_2}{Q_1+Q_2},
 \label{eq:crossslot-dq}
\end{equation}

where $Q_1$ and $Q_2$ are the portions of the left-inlet stream leaving through
the two outlets, rather than the total fluxes through those outlets.
Together, the observables record whether mirror symmetry is lost and, if so,
the direction of the labelled-inlet flow split.
The stream-function span and an independently accumulated midpoint throughput
cover the same inlet support and give the same flow split in every calculation.

The seed modifies the initial vertical velocity only at interior fluid nodes
according to

\begin{equation}
 u_y\leftarrow u_y+\epsilon U_0
 \exp\!\left[-\frac{(x-x_c)^2+(y-y_c)^2}{W^2}\right],
 \label{eq:crossslot-seed}
\end{equation}

where $(x_c,y_c)$ is the central junction and
$\epsilon\in\{-10^{-4},+10^{-4}\}$.
Open-boundary nodes retain their prescribed data.

\begin{table}[!htbp]
\centering
\caption{Window statistics for the five cross-slot calculations.}
\label{tab:crossslot-cmin}
\begingroup
\fontsize{9.5}{12}\selectfont
\renewcommand{\arraystretch}{1.18}
\setlength{\tabcolsep}{3pt}
\begin{tabular*}{170mm}{@{\extracolsep{\fill}}lrrcrrr@{}}
\toprule
Case & $\Weiss$ & Grid & $\epsilon$ & $t^*$ window & $\overline A$ & median $D_Q$ \\
\midrule
S1 & $0.55$ & W41 & $+10^{-4}$ & $351$--$400$ & $3.69\times10^{-12}$ & $5.16\times10^{-12}$ \\
S2 & $0.60$ & W41 & $+10^{-4}$ & $351$--$400$ & $5.53\times10^{-2}$ & $-1.61\times10^{-1}$ \\
S3 & $0.75$ & W41 & $-10^{-4}$ & $351$--$400$ & $2.29\times10^{-1}$ & $8.12\times10^{-1}$ \\
S4 & $0.75$ & W41 & $+10^{-4}$ & $351$--$400$ & $2.29\times10^{-1}$ & $8.12\times10^{-1}$ \\
S5 & $0.75$ & W61 & $+10^{-4}$ & $81$--$130$ & $2.33\times10^{-1}$ & $8.27\times10^{-1}$ \\
\bottomrule
\end{tabular*}
\endgroup
\par\vspace{5pt}
\begin{minipage}{170mm}
\fontsize{9}{11.5}\selectfont
\noindent
$\overline A$ and median $D_Q$ use the observables in
eqs.~\eqref{eq:crossslot-order} and \eqref{eq:crossslot-dq}.
Each reported window contains 50 samples. The common-window resolution
comparison is given separately in Table~\ref{tab:crossslot-common-window}.

\end{minipage}
\end{table}

\begin{figure}[!htbp]
\centering
\includegraphics[width=170mm]{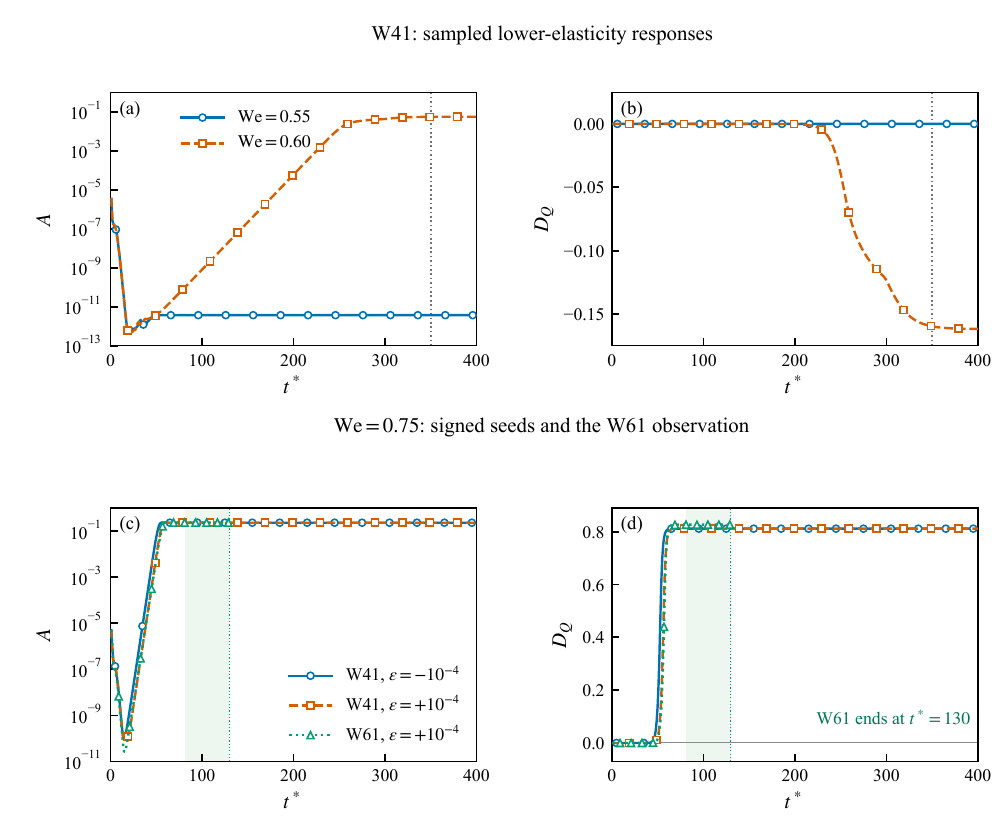}
\caption{Time histories of all five cross-slot cases.  Panels (a,b) compare
S1 and S2 on W41 at $\Weiss=0.55$ and $0.60$; panels (c,d) show S3--S5 at
$\Weiss=0.75$, with the seed signs and grids indicated in the legend.
Panels (a,c) use a logarithmic scale for the strictly positive original $A$
values, and panels (b,d) retain the sign of $D_Q$.
Lines show the available output times; symbols mark selected sampling times.
Overlapping curves are not offset.
The dotted line at $t^*=350$ precedes
the W41 statistics window $351$--$400$, whose first sample is at $t^*=351$.  Shading in (c,d) marks the common
$81$--$130$ window used for the resolution comparison.}
\label{fig:crossslot-trajectories}
\end{figure}

The W41 calculations at $\Weiss=0.55$ and $0.60$ approach qualitatively
different symmetry states, as shown by the time histories in
Figure~\ref{fig:crossslot-trajectories}(a,b).
Over the final 50 samples ($t^*=351$--400), the $\Weiss=0.55$ case has
$\overline A=3.69\times10^{-12}$ and median $D_Q=5.16\times10^{-12}$,
indicating symmetry to numerical precision.
At $\Weiss=0.60$, the corresponding values are $\overline A=5.53\times10^{-2}$
and median $D_Q=-0.161$, so a finite mirror residual accompanies a signed
imbalance in the left-inlet flow split.

\FloatBarrier

\begin{figure}[!htbp]
\centering
\includegraphics[width=170mm]{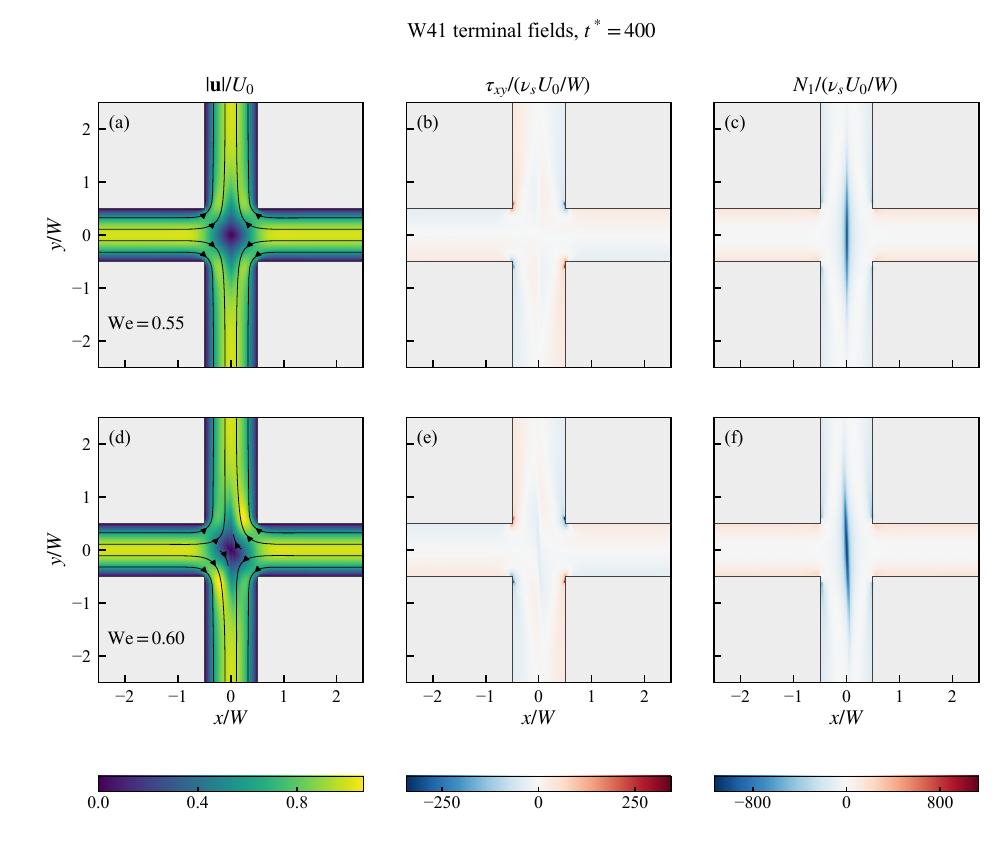}
\caption{W41 fields sampled at $t^*=400$ in the central window
$|x/W|,|y/W|\leq2.5$.  The top row is $\Weiss=0.55$ and the bottom row is
$\Weiss=0.60$; columns show speed with streamlines, shear stress, and first
normal-stress difference $N_1=\tau_{xx}-\tau_{yy}$.
Each column uses one common scale for both rows.
Velocity and stresses are normalized by $U_0$ and $\nu_sU_0/W$,
respectively; solid nodes are masked.}
\label{fig:crossslot-fields}
\end{figure}

The symmetry difference is concentrated around the central flow split
(Figure~\ref{fig:crossslot-fields}).
At $\Weiss=0.55$, the central split and the surrounding fields retain their
mirror symmetry.
At $\Weiss=0.60$, the displaced split is accompanied by asymmetric speed,
shear-stress, and first-normal-stress-difference fields.
The velocity and stress views thus place the asymmetry in the same central region.

At $\Weiss=0.75$, the opposite-seed W41 cases S3 and S4 attain the same
flow-split direction in Figure~\ref{fig:crossslot-trajectories}(c,d).
Their mean $A$ and median $D_Q$ over $t^*=351$--400 agree to the displayed
precision (Table~\ref{tab:crossslot-cmin}).
Thus, reversing this initial seed does not reverse the observed flow split
in the final W41 window.

A one-step mirror test uses a small grid with 7 cells across the channel,
14-cell arms, and $U_0=0.02$.
It compares density, velocity, all D2Q9 populations, all nine configuration
coefficients, stress, and polymer force on mirrored states.
The largest scaled difference is $5.55\times10^{-16}$, whereas an imposed
$10^{-6}$ asymmetric perturbation is detected.
This result verifies mirror equivariance for the tested one-step update.
The common flow-split direction in S3 and S4 remains unexplained.

The resolution comparison pairs positive-seed cases S4 and S5 over
$t^*=81$--130 because W41 and W61 have different final observation times.
Table~\ref{tab:crossslot-common-window} reports these common-window statistics
at $\Weiss=0.75$.
Relative to W41, the W61 mean $A$ and median $D_Q$ differ by about $1.7\%$
and $1.9\%$, respectively, using the unrounded W41 values as denominators.
The statistics and time histories show a sustained asymmetric response with
the same flow-split direction at both resolutions over this interval.

\begin{table}[!htbp]
\centering
\caption{Window statistics at $\Weiss=0.75$ for the positive-seed resolution comparison.}
\label{tab:crossslot-common-window}
\begingroup
\fontsize{9.5}{12}\selectfont
\renewcommand{\arraystretch}{1.18}
\setlength{\tabcolsep}{3pt}
\begin{tabular*}{170mm}{@{\extracolsep{\fill}}lrr@{}}
\toprule
Case & $\overline A$ & median $D_Q$ \\
\midrule
S4 (W41) & $2.29\times10^{-1}$ & $8.12\times10^{-1}$ \\
S5 (W61) & $2.33\times10^{-1}$ & $8.27\times10^{-1}$ \\
\bottomrule
\end{tabular*}
\endgroup
\par\vspace{5pt}
\begin{minipage}{170mm}
\fontsize{9}{11.5}\selectfont
\noindent
Both rows use all 50 samples in the inclusive $t^*=81$--$130$ window.
The S4 entries are evaluated over this common window; their rounded values
coincide with the W41 statistics in Table~\ref{tab:crossslot-cmin}.

\end{minipage}
\end{table}

\begin{figure}[!htbp]
\centering
\includegraphics[width=170mm]{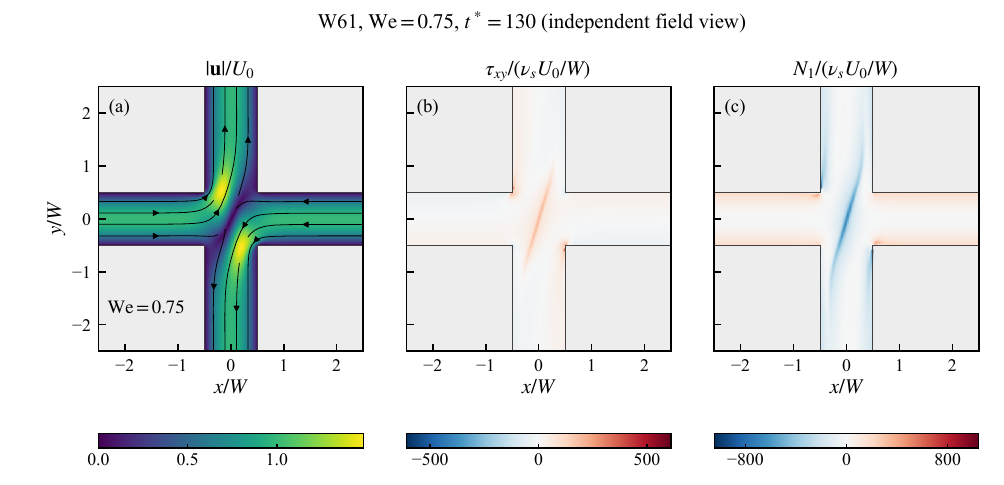}
\caption{W61 case S5 at $\Weiss=0.75$ and $t^*=130$,
in the central window $|x/W|,|y/W|\leq2.5$.
Panels (a)--(c) show speed with streamlines, shear stress $\tau_{xy}$,
and $N_1=\tau_{xx}-\tau_{yy}$, respectively.  Velocity and stresses are
normalized by $U_0$ and $\nu_sU_0/W$; solid nodes are masked.
The colour limits are set for this field view independently of
the W41 fields in Figure~\ref{fig:crossslot-fields}.}
\label{fig:crossslot-w61-fields}
\end{figure}

The sustained asymmetry on W61 is also visible in the central-junction speed, shear stress, and first normal-stress difference at $t^*=130$ (Figure~\ref{fig:crossslot-w61-fields}), accompanying the finite mirror residual and signed flow split in Table~\ref{tab:crossslot-common-window}.

\FloatBarrier

\section{Conclusions}
\label{sec:conclusions}

We have developed a deterministic Fokker--Planck/lattice-Boltzmann micro--macro solver for two-dimensional dilute Hookean polymer solutions. The configuration distribution is represented in a fixed Hermite basis and advanced through local coefficient evolution and physical-space transport. Its second moments provide the Kramers stress that feeds back into the P-TRT flow solver. Using the $M=2$ representation, we have examined the macroscopic response of this coupling in four flow configurations.

The Poiseuille calculations reproduce the analytical start-up response and steady velocity and polymer-stress profiles with both coefficient-transport discretizations. The velocity-profile error decreases along the coupled spatial and temporal refinement path. In periodic four-roll flow, comparison with an independently discretized Oldroyd--B solution tests the coupling under spatially varying extension and transport. On the $N=257$ grid, the largest relative differences across $\Weiss\in\{0.1,0.3,0.6,1.0\}$ near $t^*=10$ are $3.09\times10^{-4}$ for velocity and $3.84\times10^{-2}$ for polymer stress. The larger differences in the constitutive fields show the importance of assessing conformation and stress alongside velocity.

The cavity and cross-slot calculations extend the assessment to flows with solid walls and open boundaries. The regularized-lid cavity develops closed recirculation accompanied by a spatially nonuniform conformation field, with the strongest response near the moving lid and the upper-right wall. In the W41 cross-slot calculations, the cases at $\Weiss=0.55$ and $0.60$ approach symmetric and asymmetric states, respectively; the velocity and stress fields locate the asymmetry around the central junction. These cases demonstrate the response of the coupled solver to wall-driven circulation and open extensional flow.

Together, the analytical comparisons, independently computed macroscopic references, and boundary-flow calculations support the recovery of the tested Hookean flow and stress responses from the fixed-basis configuration dynamics. Further work will examine the effects of configuration resolution, time integration, and boundary refinement on accuracy over a wider range of flow conditions.

\section*{CRediT author contributions}
Xueqin Liu: Conceptualization (supporting), Methodology (supporting),
Software, Formal analysis, Investigation, Visualization (lead),
Writing -- original draft, Writing -- review \& editing.
Shi Shu: Supervision (supporting), Funding acquisition.
Yaolong Yu: Visualization (supporting), Writing -- review \& editing (supporting).
Yuan Yu: Conceptualization (lead), Methodology (lead), Supervision (lead),
Project administration, Writing -- review \& editing, Funding acquisition.
Hao Zhang: Visualization (supporting), Writing -- review \& editing (supporting).
Yuting Zhou: Visualization (supporting), Writing -- review \& editing (supporting).

\section*{Declaration of competing interest}
The authors declare that they have no known competing financial interests or
personal relationships that could have appeared to influence the work reported
in this paper.

\section*{Data availability}
The data that support the findings of this study are available from the
corresponding author upon reasonable request.

\section*{Declaration of generative AI and AI-assisted technologies}
During the preparation of this work, the authors used Claude (Anthropic) and
Codex (OpenAI) for language editing, bilingual consistency checks,
literature-search assistance, and audits of code, data, and plotting scripts.
The authors reviewed and edited the resulting content and take full
responsibility for the scientific decisions and the published article.
\section*{Acknowledgements}
This work was supported by the National Natural Science Foundation of China
(Grant Nos.~12101527 and 12371373), the Natural Science Foundation of Hunan
Province (No.~2026JJ60005), the Science and Technology Innovation Program of
Hunan Province (No.~2026RC3172), the 111 Project (No.~D23017), and the
Xiangtan University Postgraduate Research and Innovation Project
(No.~XDCX2026Y323).
Computational resources were provided by the High Performance Computing
Platform of Xiangtan University.
The funding sources had no involvement in the study design, in the
collection, analysis and interpretation of data, in the writing of the report,
or in the decision to submit the article for publication.

\clearpage
\appendix
\section{Method definitions and solver controls}
\label{sec:method-details}
\setcounter{equation}{0}
\renewcommand{\theequation}{\Alph{section}.\arabic{equation}}
\renewcommand{\theHequation}{method.\Alph{section}.\arabic{equation}}

\subsection{Hermite identities and low-order moments}
\label{sec:method-hermite-identities}

The polynomial identities that fix this physicists' Hermite convention and enter the local configuration discretization are
\begin{equation}
 \begin{aligned}
  \int_{-\infty}^{\infty}H_a(z)H_b(z)e^{-z^2}\,\mathrm{d}z
  &=\sqrt{\pi}\,2^a a!\,\delta_{ab},\\
  H_a'(z)&=2aH_{a-1}(z),\\
  zH_a(z)&=\tfrac12H_{a+1}(z)+aH_{a-1}(z).
 \end{aligned}
 \label{eq:method-hermite-identities}
\end{equation}

Applying these identities to eq.~\eqref{eq:method-hermite-expansion} at $\alpha=1/\sqrt2$ gives the coefficient-to-moment relation in eq.~\eqref{eq:method-coefficient-moments}. For the local update, set $\boldsymbol{G}=\nabla_h\boldsymbol{u}^{\mathrm{use}}$, held fixed over the substep. Taking the same linear combinations of the coefficient equations yields
\begin{equation}
 \frac{\boldsymbol{m}^{\mathrm{loc}}-\boldsymbol{m}^{n}}{\Delta t}
 =\boldsymbol{G}\boldsymbol{m}^{\mathrm{loc}}
  +\boldsymbol{m}^{\mathrm{loc}}\boldsymbol{G}^{\mathsf T}
  -\frac{1}{\lambda}
   \left(\boldsymbol{m}^{\mathrm{loc}}-m_0^n\boldsymbol{I}\right),
 \qquad m_0^{\mathrm{loc}}=m_0^n.
 \label{eq:method-local-moments}
\end{equation}
This is the implicit-Euler update of the local Hookean moment equation. When $m_0^n=1$, it has the conformation form of eq.~\eqref{eq:model-oldroydb} without physical-space transport. No coefficient of total degree above two enters this local subsystem.

\subsection{D2Q9 lattice and Hermite tensor}
\label{sec:method-lattice-details}

The discrete velocities and weights are
\begin{equation}
 \begin{aligned}
  \boldsymbol{c}_0&=(0,0),\\
  \boldsymbol{c}_i&=(\cos[(i-1)\pi/2],\,\sin[(i-1)\pi/2]),
    &&i=1,\ldots,4,\\
  \boldsymbol{c}_i&=\sqrt{2}\,(\cos[(2i-9)\pi/4],\,\sin[(2i-9)\pi/4]),
    &&i=5,\ldots,8,\\
  w_0&=4/9,\qquad w_{1,\ldots,4}=1/9,\qquad
  w_{5,\ldots,8}=1/36.
 \end{aligned}
 \label{eq:method-d2q9}
\end{equation}
The Hermite tensor is $H^{(3)}_{i,\mu\nu\kappa}
=c_{i\mu}c_{i\nu}c_{i\kappa}
-c_s^2(c_{i\mu}\delta_{\nu\kappa}
+c_{i\nu}\delta_{\mu\kappa}
+c_{i\kappa}\delta_{\mu\nu})$.

\subsection{Local linear solvers}
\label{sec:method-local-solvers}

Write the local coefficient system as $\boldsymbol{B}\boldsymbol{c}=\boldsymbol{b}$, where $\boldsymbol{b}$ contains the incoming coefficients and $\boldsymbol{c}$ the updated coefficients. Four-roll and Poiseuille flow use pivoted dense elimination, with AAJ as a fallback when a pivot is smaller in magnitude than $10^{-20}$ and $10^{-28}$, respectively. Cavity and cross-slot flow use AAJ directly.

AAJ combines a Jacobi fixed-point map with damping $0.8$ and up to five Anderson history vectors. If the Anderson coefficient solve fails, the damped Jacobi candidate is retained. The iteration stops when $\|\boldsymbol{c}^{k+1}-\boldsymbol{c}^{k}\|_\infty\le10^{-9}$ or the iteration count reaches $500$. The four-roll and Poiseuille fallback routines and the cavity routine return the current iterate at this limit. Cross-slot flow additionally tests the true linear-system residual and the relative zeroth-mode change:
\begin{equation}
 \frac{\|\boldsymbol{B}\boldsymbol{c}-\boldsymbol{b}\|_\infty}
 {\max(10^{-14},\|\boldsymbol{b}\|_\infty)}\le10^{-8},
 \qquad
 \frac{|c_{00}-b_{00}|}{\max(10^{-14},|b_{00}|)}\le10^{-12}.
 \label{eq:method-local-solver-checks}
\end{equation}
Failure of either cross-slot test terminates the calculation. Poiseuille and cavity restore the incoming zeroth coefficient after the local solve; cross-slot performs this restoration after its checks. Four-roll uses the zeroth-mode identity row without a separate restoration.

\clearpage

\bibliographystyle{elsarticle-num}
\bibliography{manuscript/references}

\end{document}